\documentclass{article}
\usepackage{iclr2027_conference,times}

\usepackage[utf8]{inputenc}
\usepackage[T1]{fontenc}
\usepackage{hyperref}
\usepackage{url}
\usepackage{graphicx}
\usepackage{tikz}
\usetikzlibrary{calc,arrows.meta}
\usepackage{booktabs}
\usepackage{amsfonts}
\usepackage{nicefrac}
\usepackage{microtype}
\usepackage{xcolor}

\newif\ifshowrebuttal
\showrebuttaltrue

\usepackage{etoolbox}
\usepackage{bm}
\usepackage{amsmath}
\usepackage{amsthm}
\usepackage{amssymb}
\usepackage{algorithm}
\usepackage{algorithmic}
\usepackage{subcaption}
\usepackage{booktabs}
\usepackage{multirow}
\theoremstyle{plain}
\newtheorem{theorem}{Theorem}[section]
\newtheorem{proposition}[theorem]{Proposition}
\newtheorem{lemma}[theorem]{Lemma}

\theoremstyle{definition}

\theoremstyle{remark}

\usepackage{thmtools}
\usepackage{thm-restate}

\usepackage[capitalize,noabbrev]{cleveref}

\crefname{proposition}{Proposition}{Propositions}
\Crefname{proposition}{Proposition}{Propositions}
\crefname{lemma}{Lemma}{Lemmas}
\Crefname{lemma}{Lemma}{Lemmas}
\crefname{corollary}{Corollary}{Corollaries}
\Crefname{corollary}{Corollary}{Corollaries}
\crefname{definition}{Definition}{Definitions}
\Crefname{definition}{Definition}{Definitions}
\crefname{assumption}{Assumption}{Assumptions}
\Crefname{assumption}{Assumption}{Assumptions}
\crefname{remark}{Remark}{Remarks}
\Crefname{remark}{Remark}{Remarks}

\pretocmd{\appendix}{%
  \crefalias{section}{appendix}%
  \crefalias{subsection}{subappendix}%
  \crefalias{subsubsection}{subsubappendix}%
}{}{}

\title{MATCHA: Bridging the Gap between Sample and Wall-Clock Efficiency in Model-Based Multi-Agent Reinforcement Learning}

\iclrfinalcopy

\author{
\textbf{Jianing Ye\textsuperscript{1}, Yang Zhang\textsuperscript{2},
Xiu Yuan\textsuperscript{1}, Buzhao Liu\textsuperscript{1},}\\
\textbf{David Park\textsuperscript{1}, Qihan Liu\textsuperscript{3},
Chongjie Zhang\textsuperscript{1}}\\[0.5em]
\normalfont\textsuperscript{1} Washington University in St. Louis\quad
\textsuperscript{2} Tsinghua University\quad
\textsuperscript{3} Mind Lab
}

\usepackage{thmtools}
\DeclareMathOperator*{\argmax}{arg\,max}

\begin{document}

\maketitle
\lhead{}

\begin{abstract}
  Model-based Multi-Agent Reinforcement Learning (MBMARL) has demonstrated superior sample efficiency compared to model-free baselines.
  However, its practical impact is often limited by high \emph{wall-clock} training time.
  To bridge this gap, we introduce \textbf{MATCHA}, a system-algorithm co-design that makes search-based MBMARL efficient on modern accelerators.
  At system level, MATCHA implements an end-to-end, device-centric framework that keeps data collection, planning, and learning on the accelerator, massively reducing costly CPU--GPU communication.
  At the algorithmic level, we propose \emph{Path-Guided Frontier Search (PGFS)}, which selects nodes directly from the frontier of the tree using a global score based on cumulative path statistics.
  This reduces selection and expansion to a single vectorized tensor operation, avoiding depth-dependent sequential tree traversal on the GPU.
  Across SMAX, Overcooked, and MPE, MATCHA matches or exceeds the sample efficiency of strong MBMARL baselines while substantially reducing training time, achieving up to $\mathbf{16.3\times}$ end-to-end wall-clock speedup with minimal GPU resources.

\end{abstract}

\section{Introduction}

Multi-Agent Reinforcement Learning (MARL) has enabled impressive progress on cooperative decision-making problems, but most successes still rely on model-free training that demands millions or tens of millions of environment interactions on standard benchmarks \citep{samvelyan2019smac, nguyen2020deep, hernandez2020very}. When experience is expensive, \emph{model-based} MARL (MBMARL) \citep{egorov2022scalablemultiagentmodelbasedreinforcement, wu2023modelsasagents, liu2024efficient, zhang2025decentralized, zhang2025revisitingmultiagentworldmodeling} is an appealing alternative: a learned world model \citep{ha18worldmodels, hafner2020dreamcontrollearningbehaviors, hafner2020dreamerv2, hafner2025dreamerv3} can be used to plan or to train from imagined trajectories, often reducing the number of real environment steps required to reach strong performance. Unfortunately, the practical impact of this sample efficiency is frequently negated by \emph{wall-clock} inefficiency. On standard benchmarks (e.g., SMAC \citep{samvelyan2019smac}), recent MBMARL methods can be over multiple orders of magnitude slower than model-free baselines, and training commonly stretches to multi-day runs even on modern accelerators \citep{egorov2022scalablemultiagentmodelbasedreinforcement,liu2024efficient,zhang2025decentralized}.

The sources of this latency differ across MBMARL paradigms.
Imagine-based approaches are often dominated by the cost of optimizing large models and generating long-horizon rollouts.
In this work we focus on \emph{search-based} MBMARL \citep{liu2024efficient}, where the bottleneck is more structural: \textbf{CPU-centric planning}.
Concretely, the dominant pipeline runs Monte Carlo Tree Search (MCTS) on the CPU while calling the GPU repeatedly for model inference during node expansion.
This tight iterative loop incurs large CPU--GPU communication overhead and caps throughput by available CPU cores.
As a result, adding GPUs yields limited speedup in practice, even though the underlying model inference and rollout evaluation are well-suited to accelerator hardware.

A common response is to \emph{scale out} with large distributed systems \citep{espeholt2018impalascalabledistributeddeeprl} that decompose data collection, planning, and learning into many concurrent workers.
While effective in single-agent settings, these systems are resource-intensive and difficult to reproduce in typical laboratory environments.\footnote{For example, reported performance for SpeedyZero \citep{mei2023speedyzero} uses a whole cluster of dozens of GPUs and hundreds of CPUs.}
In contrast, we pursue the complementary goal of \emph{scaling up} efficiency on minimal hardware (e.g., one or two GPUs) by addressing the root cause of latency in search-based MBMARL: the mismatch between traditional tree search and modern accelerator architectures.

Naively porting MCTS to a GPU is insufficient.
Standard MCTS executes $N$ simulations sequentially, and each simulation performs depth-$d$ pointer-chasing tree traversal plus backpropagation updates.
This creates depth-dependent sequential control flow with branch divergence that prevents SIMD (Single Instruction, Multiple Data) hardware from reaching high utilization.
Multi-agent settings further exacerbate this mismatch: the joint action space typically induces larger and deeper search trees, making conventional MCTS even more expensive in practice.

We propose \textbf{MATCHA} (\textbf{M}odel-based MARL \textbf{A}rchi\textbf{T}ecture with \textbf{C}o-designed Pat\textbf{H}-Guided Frontier Search \textbf{A}lgorithm), a system--algorithm co-design that bridges the gap between sample efficiency and wall-clock efficiency in search-based MBMARL.
MATCHA is implemented as a device-centric vectorized framework that keeps \emph{data collection, planning, and learning} resident on the accelerator, massively reducing the frequent CPU--GPU communication inherent in prior designs.
At the algorithmic level, MATCHA introduces a \textbf{Path-Guided Frontier Search (PGFS)} that replaces step-by-step tree traversal with one-shot tensor operations over the frontier.
The search maintains path-based scores for frontier nodes, and performs global batched sampling and updates across the frontier.
This removes depth-dependent sequential traversal from the search inner loop and enables massively parallel evaluation of candidate nodes on the GPU.

Our contributions are summarized as follows:
\begin{itemize}
  \item \textbf{Path-Guided Frontier Search (PGFS):} We propose a path-guided frontier search algorithm whose core operations are batched tensor primitives (batched sampling and updates), aligning with SIMD-style execution on modern accelerators.
  We also provide a theoretical analysis of the proposed search procedure and derive a non-stationary policy improvement guarantee.
  \item \textbf{Device-Centric Vectorized MBMARL framework:} Built on PGFS, we propose an end-to-end vectorized framework \textbf{MATCHA} that keeps simulation, planning, and learning on-device, massively reducing costly CPU--GPU communication in search-based MBMARL.
  \item \textbf{Practical speedups without sacrificing sample efficiency:} Extensive experiments on SMAX, Overcooked, and MPE show that MATCHA matches or exceeds the sample efficiency of strong MBMARL baselines while substantially reducing end-to-end training time, achieving up to $16.3\times$ speedup on minimal computational resources.
\end{itemize}

\section{Related Work}

\paragraph{Multi-Agent RL.}
The dominant paradigm for cooperative MARL is Centralized Training with Decentralized Execution (CTDE), which uses global state information during training while restricting execution to local observations.
Model-free methods adhering to this paradigm are broadly categorized into value-based approaches \citep{sunehag2017valuedecompositionnetworkscooperativemultiagent, rashid2018qmixmonotonicvaluefunction, son19qtran, wang2021qplexduplexduelingmultiagent} and policy-based approaches \citep{lowe2017mpe, Foerster18coma, iqbal19maac, Ryu2019MultiAgentAW, liu20maga, kuba2021trust, peng2021facmac, yu2022surprisingeffectivenessppocooperative, zhang2024read, ye2025globaloptimalitycooperativemarl}.
Despite their success, model-free algorithms suffer from high sample complexity, motivating growing interest in model-based MARL (MBMARL).
One line of work, including MAMBA \citep{egorov2022scalablemultiagentmodelbasedreinforcement}, MARIE \citep{zhang2025decentralized}, and DIMA \citep{zhang2025revisitingmultiagentworldmodeling}, adapts Dreamer \citep{hafner2020dreamcontrollearningbehaviors} to multi-agent settings, leveraging learned world models for policy optimization via latent imagination.
Another line focuses on search-based planning; notably, MAZero \citep{liu2024efficient} extends MuZero \citep{Schrittwieser_2020} to enable multi-agent policy improvement via Monte-Carlo Tree Search (MCTS).
While these model-based methods improve sample efficiency over model-free baselines, they incur substantial computational cost, especially search-based methods that rely on sequential planning and CPU-centric implementations.

\paragraph{Scalable and Distributed RL.}
Acceleration strategies generally fall into two categories: \textit{scale-out} and \textit{scale-up}.
Scale-out approaches rely on massive distributed infrastructure.
Model-free systems \citep{nair2015massivelyparallelmethodsdeep,horgan2018distributedprioritizedexperiencereplay,espeholt2018impalascalabledistributeddeeprl} and model-based systems \citep{ye2021masteringatarigameslimited,wang2024efficientzerov2masteringdiscrete,mei2023speedyzero} have achieved high speeds via asynchronous execution on large clusters, but their resource demands limit accessibility and often increase sensitivity to hyperparameter tuning.
Scale-up approaches focus on algorithmic efficiency.
Techniques like Sampled MuZero \citep{hubert2021learningplanningcomplexaction}, Gumbel MuZero \citep{danihelka2022policy}, and V-MCTS \citep{ye2022spendingthinkingtimewisely} reduce search costs by optimizing simulation counts or action spaces.
However, they retain the sequential execution and tree-traversal structure of MCTS, and thus cannot fully exploit the massive parallelism of modern GPUs.

\paragraph{Vectorized and Hardware-Accelerated RL.}
The rise of JAX has enabled end-to-end vectorized environments \citep{gymnax2022github, koyamada2023pgx, radji2026octax}.
Libraries like Brax \citep{brax2021github}, Jumanji \citep{bonnet2024jumanji} and JaxMARL \citep{flair2024jaxmarl} provide comprehensive benchmark suites, including MARL environments SMAX, Overcooked, and MPE.
By leveraging massive parallelism and on-device execution, these environments make large-scale parallel simulation on GPUs possible.
Building on this foundation, efficient end-to-end vectorized implementations have been established for model-free algorithms \citep{jaxrl,flair2024jaxmarl, hessel2021podracerarchitecturesscalablereinforcement, dekock2023mavaresearchlibrarydistributed, mahjoub2025sableperformantefficientscalable}.
However, efficient vectorized implementations for search-based MBMARL remain absent.
Existing related efforts \citep{toledo2024stoix} are limited to single-agent settings and do not provide a complete and efficient realization of search-based MBRL in a device-centric framework.
To the best of our knowledge, our work is the first to close this gap for search-based MBMARL.

\section{Preliminaries}

\subsection{Search-Based MBRL and Reanalysis}

Search-based reinforcement learning methods such as MuZero and its multi-agent extensions typically follow a three-stage training pipeline:
(1) \textbf{Data Collection}, where agents interact with the environment using a search-based policy
and store trajectories in a replay buffer;
(2) \textbf{Reanalysis}, where historical states are sampled from the buffer and re-evaluated
using a fresh search to produce improved learning targets; and
(3) \textbf{Optimization}, where model parameters are updated to fit these targets.

Among these stages, reanalysis is essential for sample efficiency and training stability but is computationally expensive, as it requires executing planning algorithms over large batches of states.
To reduce the latency of the reanalysis process, previous MBRL algorithms usually launch multiple reanalysis workers on host-side to reanalyze the data in a concurrent and asynchronous manner.

\subsection{Monte Carlo Tree Search Variants}

Monte Carlo Tree Search (MCTS) and its variants \citep{hubert2021learningplanningcomplexaction,danihelka2022policy,liu2024efficient,wang2024efficientzerov2masteringdiscrete} build a search tree rooted at an initial state $s_0$ by iteratively performing three phases:
\textbf{Selection}, \textbf{Expansion}, and \textbf{Backpropagation}.
Each node in the tree corresponds to a state, and each edge corresponds to an action.
During selection, the algorithm traverses the tree from the root to a leaf according to stored statistics such as action priors and value estimates.
Upon reaching a leaf, the learned model is used to expand new child nodes and to estimate their values.
Backpropagation then updates statistics along the traversed path, which influences subsequent selections.

\section{Method: MATCHA Framework}
\label{sec:method}

In this section, we introduce \textbf{MATCHA} (\textbf{M}odel-based MARL \textbf{A}rchi\textbf{T}ecture with \textbf{C}o-designed Pat\textbf{H}-Guided Frontier Search \textbf{A}lgorithm), a high-throughput MBMARL framework designed to bridge the gap between sample efficiency and time efficiency.
MATCHA achieves this through a system-algorithm co-design:
(1) A device-centric end-to-end vectorized architecture for MBMARL that confines the entire training lifecycle within accelerator memory;
(2) The Path-Guided Frontier Search (PGFS) algorithm, a fundamentally different search paradigm that replaces the sequential tree traversal of traditional MCTS variants with one-shot tensor operations.
We will first detail the device-centric architecture and then introduce the PGFS algorithm.

\subsection{Device-Centric End-to-End Vectorized Architecture}
\label{subsec:arch}

\begin{table}[t]
\centering
\scriptsize
\setlength{\tabcolsep}{2.3pt}
\caption{Per-component profiling results of different algorithms on two SMAX environments with two RTX 5090. Hyper-parameters are aligned with the MAZero baseline for fair comparison.}
\begin{tabular}{l|ccc|ccc}
\toprule
& \multicolumn{3}{c|}{\textbf{3s\_vs\_5z} (sec/train step) $\downarrow$} 
& \multicolumn{3}{c}{\textbf{2s3z} (sec/train step) $\downarrow$} \\
\textbf{Algorithm} 
& \textbf{Data} & \textbf{Reanalysis} & \textbf{Optim.} 
& \textbf{Data} & \textbf{Reanalysis} & \textbf{Optim.} \\
\midrule
MATCHA & \textbf{0.0010} & \textbf{0.0045} & \textbf{0.0092} &
\textbf{0.0010} & \textbf{0.0067} & \textbf{0.0091} \\
MATCHA w/ GPU MCTS & 0.0200 & 0.0585 & \textbf{0.0123} &
0.0220 & 0.0548 & \textbf{0.0122} \\
MATCHA w/ GPU Gumbel MCTS & 0.0157 & 0.0261 & \textbf{0.0113} &
0.0159 & 0.0260 & \textbf{0.0112} \\
MAZero & 0.1712 & 0.1472 & 0.0838 &
0.1720 & 0.1236 & 0.0889 \\
\bottomrule
\end{tabular}

\label{tab:profiling}
\end{table}

\begin{figure*}[t]
\centering
\includegraphics[width=0.8\textwidth]{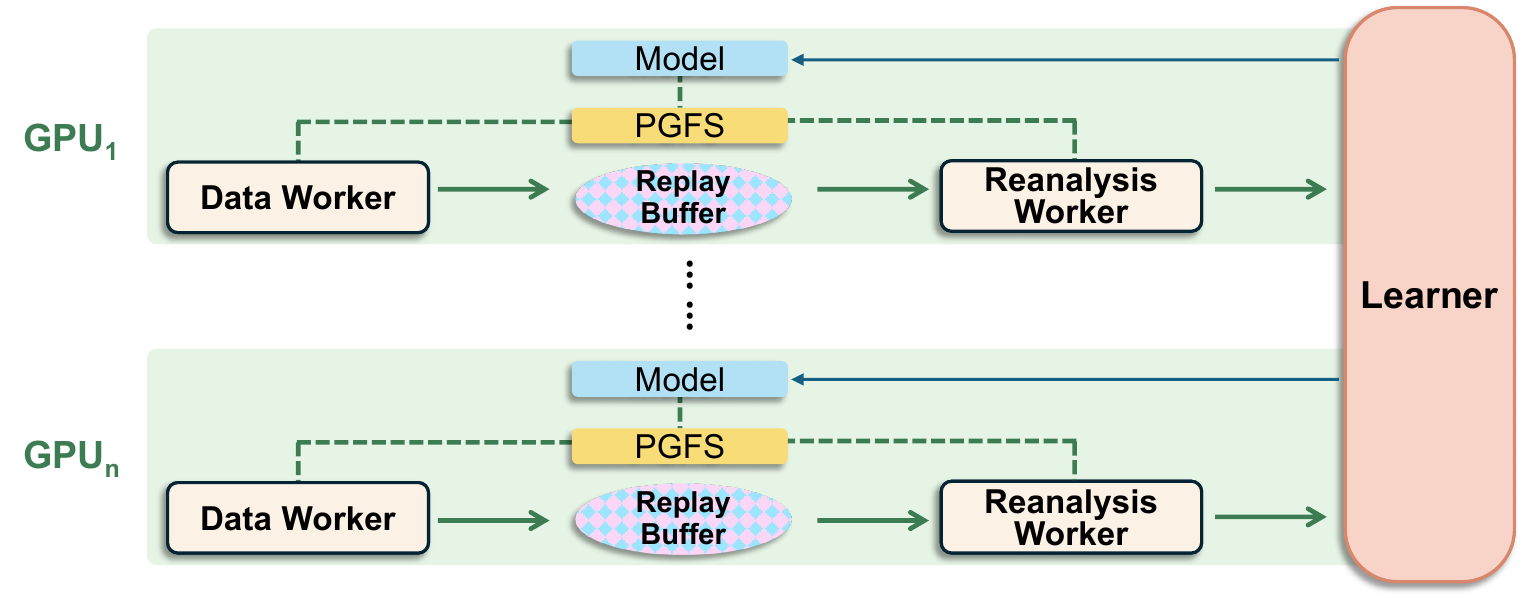}
\caption{Overview of MATCHA's device-centric, end-to-end vectorized architecture. Each accelerator hosts a self-contained processing pipeline, while a centralized learner coordinates learning across devices. The patterned replay buffer denotes flexible buffer placement: GPU-resident by default, with CPU-resident buffers supported for high-dimensional observation spaces such as image inputs.}
\label{fig:MATCHA-architecture}
\end{figure*}

In the standard CPU-centric training lifecycle of search-based MBMARL, illustrated by MAZero in \cref{fig:MAZero-architecture}, the two dominant computational bottlenecks are \textbf{Data Collection} and \textbf{Reanalysis} (profiled in \cref{tab:profiling}), as they require executing expensive planning algorithms at every step.
Reanalysis, in particular, is computationally prohibitive because it requires running MCTS on large batches of data simultaneously for sample efficiency and training stability.
Directly parallelizing this batched search on standard hardware is non-trivial due to two system-level constraints:
(1) the degree of parallelism is hard-capped by the number of available CPU cores, leaving the GPU unsaturated;
(2) the iterative nature of MCTS necessitates frequent bidirectional data transfer between the CPU (search logic) and GPU (inference), creating a severe host-device communication bandwidth bottleneck that cannot be resolved simply by adding more threads. MATCHA addresses these constraints with a \textbf{Device-Centric End-to-End Vectorized Architecture} that performs storage and computation directly on the accelerator whenever possible (see \cref{fig:MATCHA-architecture}). 

\paragraph{On-device reanalysis chains.}
On each of the $n$ available accelerators, MATCHA instantiates a self-contained on-device chain for batched data collection, planning, reanalysis, and model inference.
Each chain runs independently and in parallel, producing a reanalyzed batch while keeping the search logic and model computation on the same device.
The replay buffer in this chain is GPU-resident by default to avoid host-device transfers, but MATCHA also supports CPU-resident buffers for high-dimensional observations such as images.
Since buffer access is sparse relative to planning and model inference, this fallback does not dominate the runtime in our setting.

\paragraph{Centralized learner.}
We use a centralized learner to aggregate the cross-device reanalyzed batches produced by each chain, perform model updates for them sequentially, and then broadcast the updated parameters back to the on-device chains for the next iteration.\footnote{In practice, MATCHA replaces explicit parameter broadcasting with implicit parameter synchronization. We place a fully synchronized learner on each device and ensure that all learners use exactly the same data and perform identical computations, thereby keeping model parameters synchronized across devices.}
This design parallelizes the expensive reanalysis stage across accelerators while keeping the learning process centralized and synchronized.

Unlike Distributed Data Parallel (DDP)-style frameworks such as Anakin \citep{hessel2021podracerarchitecturesscalablereinforcement}, which improve environment-step throughput by forming a larger effective batch across devices and synchronizing local updates, MATCHA consumes the $n$ reanalyzed batches as $n$  sequential optimization updates.
This enables MATCHA to reduce the average latency per optimization step without reducing the Update-to-Data ratio (UTD), which is important for preserving sample efficiency \citep{doro2023sampleefficient}.
In contrast, DDP-style scaling can linearly reduce the effective UTD ratio as the number of devices increases, potentially degrading sample efficiency.
We provide a detailed ablation study for UTD in \cref{appsubsec:UTD}, and a more detailed comparison with DDP-style scaling in \cref{appsubsec:DDP}.

Together, the on-device reanalysis chains and centralized learner substantially reduce CPU--GPU communication overhead while effectively leveraging the parallelism of modern accelerators.

\subsection{Path-Guided Frontier Search}

For traditional search algorithms such as MCTS and its variants, merely porting them to the device is insufficient to unleash the full potential of modern accelerators (\cref{fig:selection-scalability}).
The core reason is that the serial tree traversal procedure in selection and backpropagation phases is not suitable for SIMD devices like GPU.
This procedure involves sequential pointer chasing and branch-heavy control flow whose cost scales with the search depth $d$, preventing the hardware from utilizing its massive parallel throughput.
This motivates a new search procedure that eliminates the sequential tree traversal from selection and backpropagation.

\definecolor{treegray}{RGB}{174,183,194}
\definecolor{nodegray}{RGB}{119,130,145}
\definecolor{mctsblue}{RGB}{48,113,181}
\definecolor{mctsfill}{RGB}{220,234,249}
\definecolor{pgfsgreen}{RGB}{38,139,104}
\definecolor{pgfsfill}{RGB}{221,241,232}
\definecolor{selectedred}{RGB}{206,70,54}
\definecolor{selectedfill}{RGB}{252,229,224}

\tikzset{
  edge/.style={draw=treegray, line width=0.9pt, line cap=round},
  node/.style={circle, draw=nodegray, fill=white, line width=0.8pt, minimum size=5.5mm, inner sep=0pt},
  mpath/.style={draw=mctsblue, line width=2.0pt, line cap=round},
  mstep/.style={mpath, -{Latex[length=2.1mm,width=1.5mm]}, shorten <=3.2mm, shorten >=3.2mm},
  mnode/.style={circle, draw=mctsblue, fill=mctsfill, line width=1.2pt, minimum size=5.8mm, inner sep=0pt},
  fnode/.style={circle, draw=pgfsgreen, fill=pgfsfill, line width=1.1pt, minimum size=5.6mm, inner sep=0pt},
  selected/.style={circle, draw=selectedred, fill=selectedfill, line width=1.5pt, minimum size=6.2mm, inner sep=0pt},
  frontiercut/.style={draw=pgfsgreen, line width=.8pt, dash pattern=on 2.4pt off 2.4pt, line cap=butt},
}

\newcommand{\drawtree}{%
  \draw[edge] (r) -- (a) (r) -- (c)
              (a) -- (lone) (a) -- (ltwo) (ltwo) -- (lseven) (ltwo) -- (leight)
              (c) -- (d) (c) -- (lfour) (d) -- (lfive) (d) -- (lsix);
  \foreach \n in {r,a,c,d,lone,ltwo,lseven,leight,lfour,lfive,lsix}
    \node[node] (n-\n) at (\n) {};
}

\begin{figure}[ht]
    \centering

    \begin{subfigure}{0.48\linewidth}
        \centering
        \raisebox{0.33cm}{\resizebox{\linewidth}{!}{%
        \begin{tikzpicture}[font=\sffamily]
          \path[use as bounding box] (-0.4,0.75) rectangle (13.95,-7.05);
          \node[font=\bfseries\Large, align=center] at (2.75,0.35) {MCTS Tree Traversal\\Selection};
          \node[font=\bfseries\Large, align=center] at (10.25,0.35) {PGFS Frontier Sampling\\Selection};
          \draw[black!18, line width=.5pt] (6.5,-0.45) -- (6.5,-6.05);

          \begin{scope}[shift={(0,-1.1)}]
            \begin{scope}[xshift=.45cm]
            \coordinate (r) at (2.3,0);
            \coordinate (a) at (1.10,-1.25);
            \coordinate (c) at (3.50,-1.25);
            \coordinate (lone) at (0.30,-2.50);
            \coordinate (ltwo) at (1.50,-2.50);
            \coordinate (d) at (3.10,-2.50);
            \coordinate (lfour) at (4.30,-2.50);
            \coordinate (lseven) at (1.10,-3.75);
            \coordinate (leight) at (1.90,-3.75);
            \coordinate (lfive) at (2.70,-3.75);
            \coordinate (lsix) at (3.50,-3.75);
            \drawtree
            \draw[mstep] (r) -- (c);
            \draw[mstep] (c) -- (d);
            \draw[mstep] (d) -- (lsix);
            \foreach \n in {r,c,d} \node[mnode] at (\n) {};
            \node[selected] at (lsix) {};
            \node[font=\normalsize\bfseries, text=mctsblue, align=left, anchor=west] at (3.35,-0.52) {tree traversal\\path};
            \node[font=\normalsize, text=selectedred, anchor=north] at (3.50,-4.08) {selected leaf $x^\star$};
            \end{scope}
            \node[font=\Large, text=black, align=center] at (2.75,-5.35) {serial iterative\\local selection};
          \end{scope}

          \begin{scope}[shift={(7.4,-1.1)}]
            \begin{scope}[xshift=.55cm]
            \coordinate (r) at (2.3,0);
            \coordinate (a) at (1.10,-1.25);
            \coordinate (c) at (3.50,-1.25);
            \coordinate (lone) at (0.30,-2.50);
            \coordinate (ltwo) at (1.50,-2.50);
            \coordinate (d) at (3.10,-2.50);
            \coordinate (lfour) at (4.30,-2.50);
            \coordinate (lseven) at (1.10,-3.75);
            \coordinate (leight) at (1.90,-3.75);
            \coordinate (lfive) at (2.70,-3.75);
            \coordinate (lsix) at (3.50,-3.75);
            \drawtree
            \foreach \n in {lone,leight,lfour,lfive,lsix} \node[fnode] at (\n) {};
            \node[selected] at (lseven) {};
            \draw[frontiercut] (-0.15,-2.10) -- (0.90,-2.10);
            \draw[frontiercut] (0.90,-2.10) -- (0.90,-3.10);
            \draw[frontiercut] (0.90,-3.10) -- (3.70,-3.10);
            \draw[frontiercut] (3.70,-3.10) -- (3.70,-2.10);
            \draw[frontiercut] (3.70,-2.10) -- (4.75,-2.10);
            \node[font=\normalsize\bfseries, text=pgfsgreen, anchor=north] at (4.70,-2.72) {frontier $\mathcal F$};
            \node[font=\normalsize, text=selectedred, anchor=north] at (1.10,-4.08) {selected leaf $x^\star$};
            \end{scope}
            \node[font=\Large, text=black, align=center] at (2.85,-5.35) {parallelizable one-shot\\global selection};
          \end{scope}
        \end{tikzpicture}%
        }}
        \caption{MCTS vs.\ PGFS selection.}
        \label{fig:selection-comparison-diagram}
    \end{subfigure}
    \hfill
    \begin{subfigure}{0.48\linewidth}
        \centering
        \includegraphics[width=\linewidth]{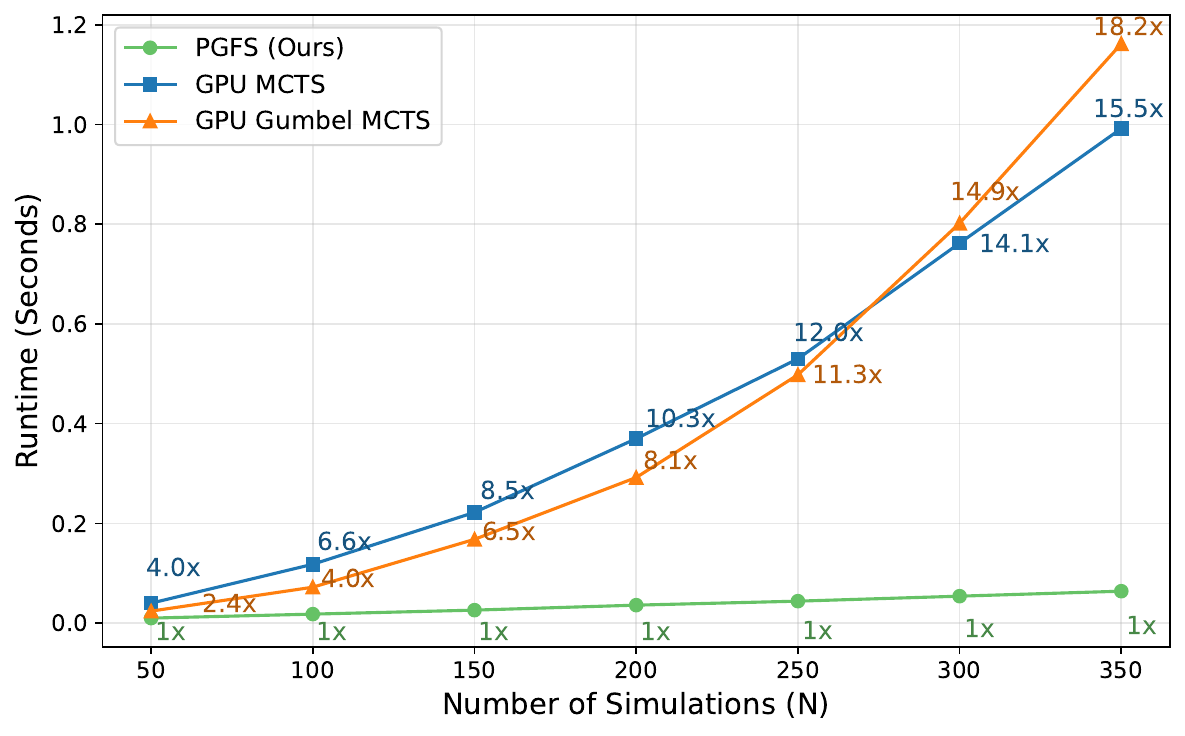}
        \caption{Scalability of search algorithms on GPU.}
        \label{fig:search-scalability-k10}
    \end{subfigure}
    \caption{
    \textbf{Search algorithm comparison.}
    (a) MCTS selects a leaf through a serial, iterative sequence of local comparisons along a single root-to-leaf traversal path (highlighted in blue), whereas PGFS scores every leaf in the frontier $\mathcal F$ in one parallel, global operation and samples directly from it.
    (b) Wall-clock runtime for one complete search under different simulation budgets $N$; annotations show runtime multipliers relative to PGFS.
    }
    \label{fig:selection-scalability}
\end{figure}

\subsubsection{Search as Frontier Sampling}
Traditional MCTS requires traversal because selection decisions are made \textit{locally}, node by node, using \textit{subtree} statistics such as value estimates and visit counts.
After that, backpropagation is also required to update the subtree information.

The key idea behind PGFS is to conduct a \textit{global selection} with a one-shot frontier sampling, by maintaining the statistics of the \textit{root-to-node path} for each node in the frontier $\mathcal F$ (the set of unexpanded leaf nodes); see \cref{fig:selection-comparison-diagram} for an illustration of this contrast.
This also removes the need for recursive backpropagation, since the cumulative path statistics of newly expanded children can be computed directly from their parent.

Formally, for every node $x$ in the frontier $\mathcal F$, we maintain three cumulative statistics of the root-to-$x$ path:
1) $\bm {l(x)}$: the log prior probability of the path;
2) $\bm {g(x)}$: the cumulative discounted reward of the path;
and 3) $\bm {d(x)}$: the product of discount factors along the path.

For any node $x \in \mathcal{F}$, we define its path-based value score $V(x)$ as:
\begin{equation}
V(x) = g(x) + d(x) \cdot v_\theta(x)
\end{equation}
which is the sum of the cumulative reward and the discounted state-value (reward-to-go).
Both rewards and state-value are estimated by the learned model.
This estimates the expected return of selecting leaf $x$.

With these statistics, the Selection and Expansion are formulated with vectorized operations:
\begin{enumerate}
    \item \textbf{Vectorized Selection}: Instead of walking the tree, we directly sample a node $x^*$ from $\mathcal{F}$ based on a Boltzmann distribution derived from our cumulative statistics:
    \begin{align}
        \eta(x)& \propto \exp\left( l(x) + \alpha \cdot V(x) \right) \label{eq:action_selection}
    \end{align}
    where $\alpha$ is a temperature hyperparameter balancing priors and value scores.
    This reduces selection to sampling over a flattened frontier tensor.

    \item \textbf{Vectorized Expansion}: For the selected node $x^*$, we invoke the model to expand it, by computing the prior probability, immediate reward, and the reward-to-go for all its children $y$.
    The new children are then added to $\mathcal{F}$, and their cumulative statistics $l(\cdot), g(\cdot), d(\cdot)$ are updated via simple vector addition and multiplication.
    After that, $x^*$ is removed from $\mathcal F$.
\end{enumerate}

The complete procedure is summarized in \cref{alg:PGFS_theoretical}.

\subsubsection{Theoretical Properties}
\label{subsubsec:theory}

In this section, we characterize the theoretical properties of PGFS (\cref{alg:PGFS_theoretical}) from three perspectives: node selection as trust-region value maximization, policy improvement, and time complexity.

\paragraph{Node Selection as Trust-Region Value Maximization}
The PGFS node-selection rule in \cref{eq:action_selection} can be viewed as a trust-region value maximization.
Specifically, it selects a node in the frontier that maximizes expected path-based value while remaining in-distribution with respect to the prior.
We characterize this property with the following proposition, which is proved in \cref{appsubsec:proof}.

\begin{restatable}{proposition}{LeafSelectionThm}
\label[proposition]{thm:leafselection}
The node selection distribution $\eta$ defined in \cref{eq:action_selection} is the optimal solution of the following constrained optimization problem:
$$
\begin{aligned}
\argmax_{\mu\in \Delta(\mathcal F)} \quad & \sum_{x\in\mathcal F}\mu(x)V(x) \\
\text{s.t.} \quad & \mathrm{KL}(\mu\|P_{\mathcal F})\le \epsilon
\end{aligned}
$$
    where $P_{\mathcal F}(x) = \exp(l(x))$ is the prior probability of the root-to-$x$ path, defined for all nodes in the frontier, $\epsilon$ is defined by \cref{DV-eps-def}.
\end{restatable}

\paragraph{Policy Improvement} MATCHA can be viewed as an instance of generalized policy iteration, in which PGFS acts as a policy improvement operator.
The following proposition establishes a policy improvement guarantee for the non-stationary policy induced by PGFS.
Its formal statement and proof are deferred in \cref{appsubsec:proof}.

\begin{proposition}[Policy Improvement, informal]
\label[proposition]{thm-informal:PI}
    If $v_\theta$ is the value function of the prior policy $\pi$, then the non-stationary policy $\widetilde\pi$ induced by PGFS achieves an expected return no lower than that of the prior policy $\pi$.
\end{proposition}

\paragraph{Time Complexity}
By removing the dependency on tree depth $d$, PGFS reduces the worst-case search complexity from $O(N^2)$ to $O(N\log N)$, where $N$ is the number of simulations (see \cref{fig:search-scalability-k10}).
We conduct a detailed analysis for the time complexity of the search algorithms in \cref{appsubsec:timecomplexityanalysis}.

\section{Experiments}
\label{sec:experiments}

In this section, we conduct experiments in three aspects:
\textbf{(A1) Wall-Clock Efficiency:} How much does MATCHA accelerate training relative to standard CPU-centric search-based MBMARL baselines?
\textbf{(A2) Sample Efficiency:} Does MATCHA retain the sample-efficiency advantage of strong model-based MARL baselines despite replacing standard tree search?
\textbf{(A3) Ablation Study:} How do PGFS and the device-centric architecture contribute to the wall-clock efficiency separately?
Is MATCHA sensitive to its hyper-parameter selections?

\subsection{Experimental Setup}
\paragraph{Environments.}
We evaluate MATCHA on three cooperative MARL benchmarks implemented in fully vectorized JAX environments:
(1) \textbf{SMAX} \citep{flair2024jaxmarl}, a hardware-accelerated SMAC-like benchmark covering a diverse set of combat scenarios; we evaluate on 10 scenarios spanning both SMACv1-like and SMACv2-like settings.
(2) \textbf{Overcooked} \citep{carroll2019overcooked, flair2024jaxmarl}, a coordination-intensive benchmark with challenging exploration under tight time constraints.
(3) \textbf{MPE} \citep{lowe2017mpe}, a suite of partially observable cooperative tasks in a 2D physical world with communication.

\paragraph{Baselines.}
We compare MATCHA against two categories of strong baselines.
(1) \textbf{Model-based baselines:} MAZero \citep{liu2024efficient} as a representative search-based method, and MAMBA \citep{egorov2022scalablemultiagentmodelbasedreinforcement} as a representative imagine-based method.
(2) \textbf{Model-free baselines:} the policy-based methods MAPPO and IPPO \citep{yu2022surprisingeffectivenessppocooperative}, and the value-based methods QMIX \citep{rashid2018qmixmonotonicvaluefunction}, VDN \citep{sunehag2017valuedecompositionnetworkscooperativemultiagent}, and IQL.
For plotting, all methods are evaluated with 5 random seeds.

\begin{figure*}[htb]
    \centering
    \includegraphics[width=\textwidth]{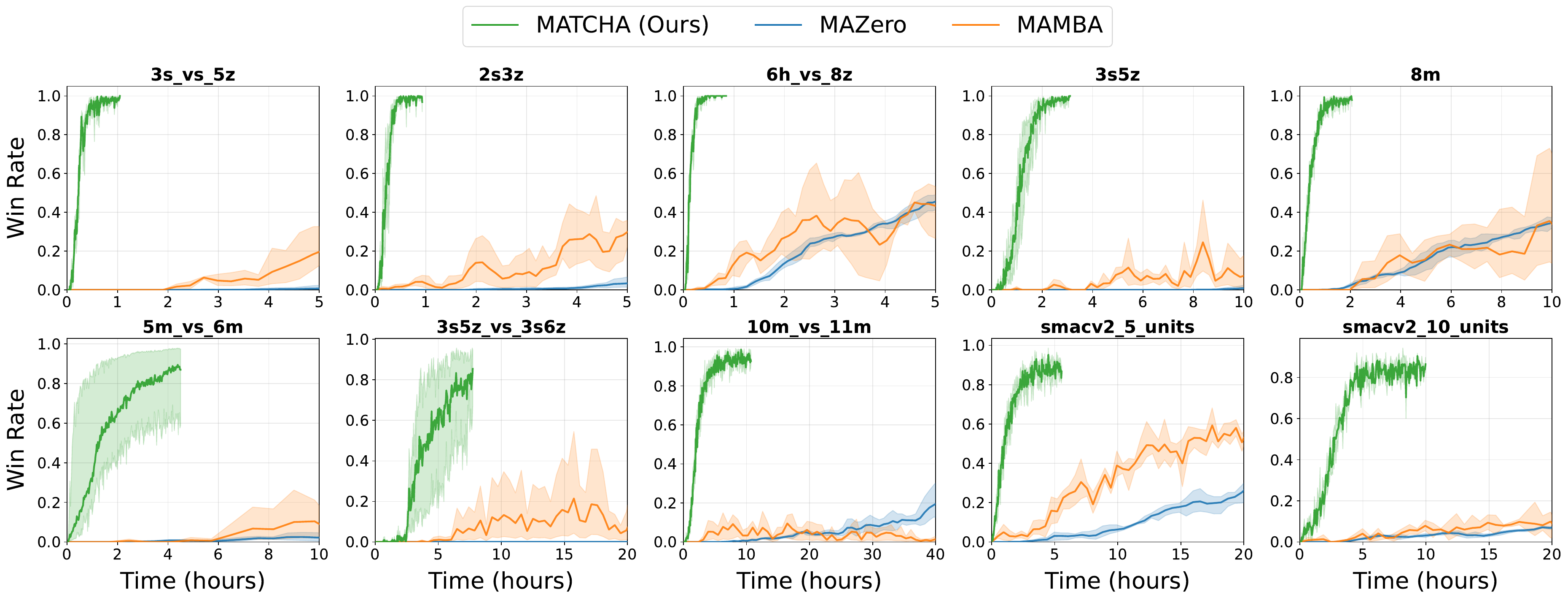}
    \caption{\textbf{Wall-clock efficiency (win rate vs. time) on SMAX.} The results are evaluated using 1 RTX 5090 GPU. See \cref{app:wall-clock} for the full results and details of the resource settings.}
    \label{fig:smax_time}
\end{figure*}

\begin{figure*}[htb]
    \centering
    \includegraphics[width=\textwidth]{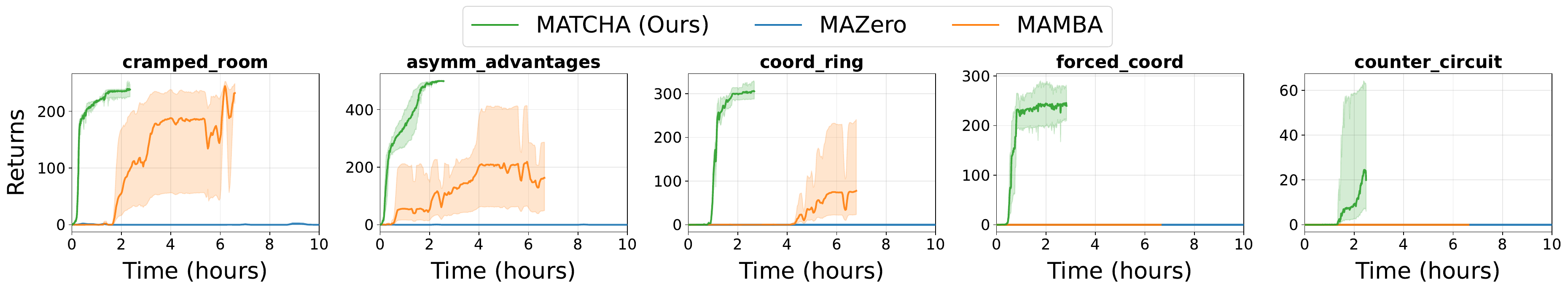}
    \caption{\textbf{Wall-clock efficiency (returns vs. time) on Overcooked.} The results are evaluated using 1 RTX 5090 GPU. See \cref{app:wall-clock} for the full results and details of the resource settings.}
    \label{fig:overcooked_time}
\end{figure*}

\begin{table}[ht]
\centering
\caption{\textbf{Wall-Clock Speed Benchmarking.} Comparison of training throughput averaged across SMAX and Overcooked tasks. MATCHA achieves an order-of-magnitude speedup over the baseline.}
\label{tab:speed_comparison}
\scriptsize
\begin{sc}
\begin{tabular}{lcccc}
\toprule
& \multicolumn{3}{c}{Training Throughput (FPS) $\uparrow$} & \\
\cmidrule(lr){2-4}
Method
& SMAX
& Overcooked
& Average
& Speedup \\
\midrule
MAZero (Baseline) & 3.54 & 7.11 & 4.73 & 1.00$\times$ \\
\textbf{MATCHA (Ours)} & \textbf{63.84} & \textbf{103.42} & \textbf{77.03} & \textbf{16.30$\times$} \\
\bottomrule
\end{tabular}
\end{sc}
\end{table}

\subsection{Wall-Clock Efficiency (A1)}

\cref{fig:smax_time,fig:overcooked_time} show the performance as a function of wall-clock time on SMAX and Overcooked, demonstrating MATCHA's ability to approach asymptotic performance much faster than the baselines.
To quantify this acceleration, \cref{tab:speed_comparison} provides a direct comparison of training throughput (FPS) averaged across SMAX and Overcooked tasks.
The quantitative results are consistent with the learning curves: MATCHA achieves a \textbf{16.3$\times$} overall speedup compared to the standard CPU-centric baseline MAZero.
These findings support our central claim that the dominant bottlenecks in MBMARL stem from CPU--GPU communication and sequential search—both of which are effectively eliminated by our design.

In addition, we conduct further experiments on MPE to broaden the diversity of our benchmark suite (\cref{app:mpe_results}), as well as an 8-GPU scalability study to verify that MATCHA continues to scale as more GPUs become available (\cref{appsubsec:gpu-scalability}).
Although the experiments in this paper focus on GPU environments to present a clean and convenient comparison, MATCHA is not limited to GPU environments and is equally effective on CPU-centric ones, as shown in \cref{appsubsec:smac-cpu}.

\subsection{Sample Efficiency (A2)}

\begin{figure*}[htb]
    \centering
    \includegraphics[width=\textwidth]{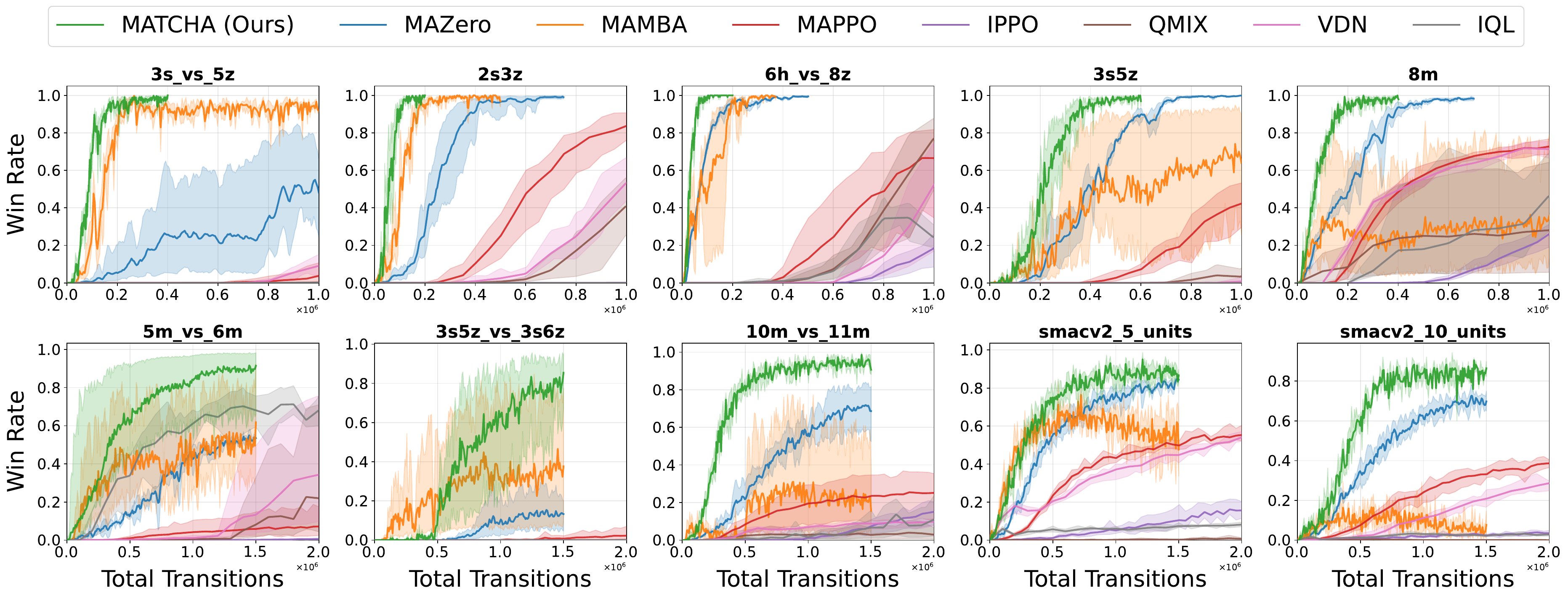}
    \caption{\textbf{Sample efficiency (win rate vs. total transitions) on SMAX.}}
    \label{fig:smax_results}
\end{figure*}

\begin{figure*}[htb]
    \centering
    \includegraphics[width=0.95\textwidth]{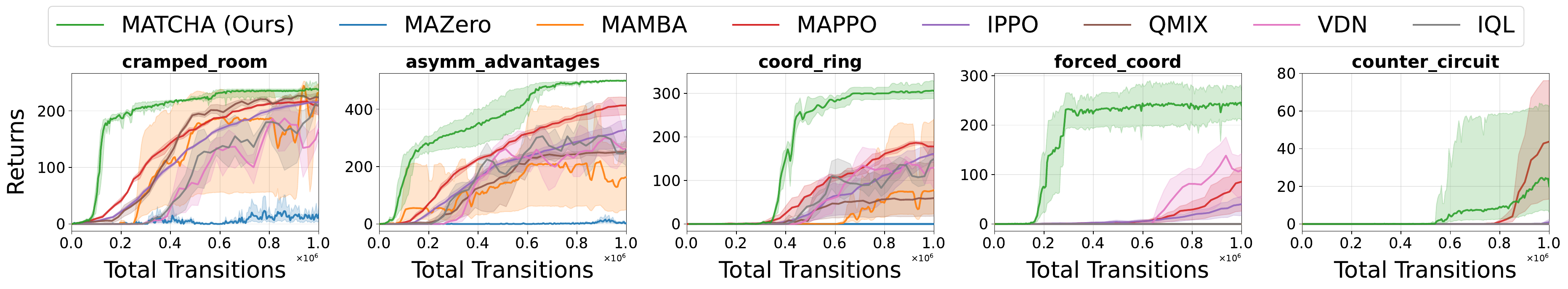}
    \caption{\textbf{Sample efficiency (returns vs. total transitions) on Overcooked.}}
    \label{fig:overcooked_results}
\end{figure*}

As shown in \cref{fig:smax_results,fig:overcooked_results,fig:mpe_results_sample_efficiency}, MATCHA consistently achieves the highest sample efficiency among all compared baselines across SMAX, Overcooked, and MPE.
These significant gains over model-free methods show that MATCHA preserves the core sample-efficiency advantage of model-based multi-agent RL.
Notably, MATCHA also outperforms strong model-based baselines, with especially large margins on Overcooked, where both MAZero and MAMBA perform poorly.
We hypothesize that stronger exploration in PGFS may partly explain these gains.
The \textit{global selection} mechanism in PGFS may improve exploration relative to standard MCTS-style \textit{iterative local selection} by comparing all frontier branches directly, rather than filtering them iteratively through a sequence of local subtree comparisons.
As a result, a promising branch may be less likely to be overlooked merely because one of its ancestors has uncompetitive subtree statistics caused by other weak sibling branches.
This interpretation is consistent with the poor performance of MAZero and MAMBA on the sparse-reward, long-horizon Overcooked tasks, and is also consistent with the ablation in \cref{fig:ablation_search_time}, where MATCHA with PGFS matches MATCHA with GPU Gumbel MCTS, which in turn outperforms MATCHA with standard GPU MCTS.

\subsection{Ablation Study and Per-component Profiling (A3)}
\label{sec:ablation}

\begin{figure}[ht]
    \centering
    \includegraphics[width=0.95\linewidth]{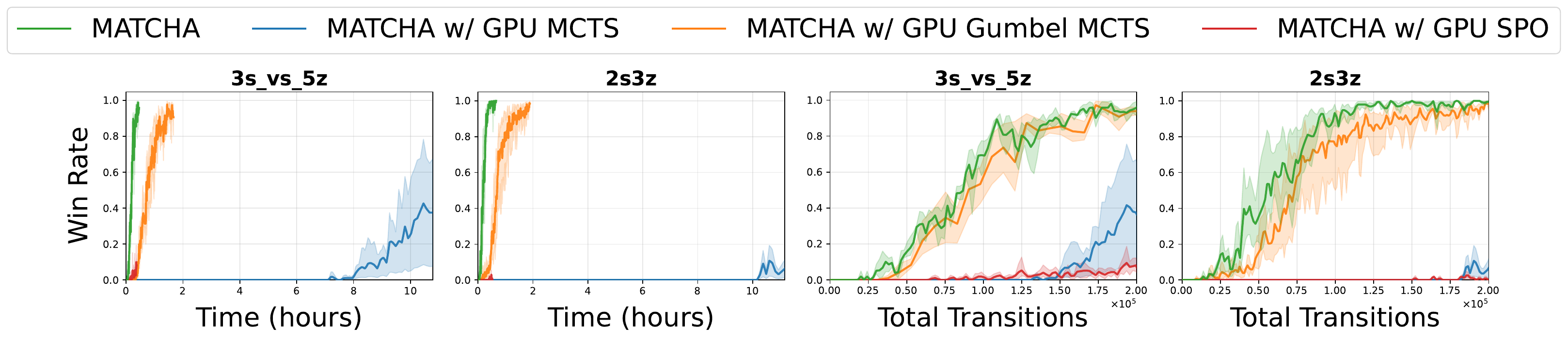}
    \caption{\textbf{Search Algorithm Comparison.} We compare MATCHA with PGFS (ours) against MATCHA with GPU-resident implementations of MCTS and Gumbel MCTS in both sample and wall-time efficiency.}
    \label{fig:ablation_search_time}
\end{figure}

We also perform an ablation study to disentangle the contributions of the system architecture and the algorithmic design.

\paragraph{Search Algorithm Comparison.}
To verify that PGFS improves wall-clock efficiency over MCTS without sacrificing sample efficiency, we compare MATCHA against three variants in the same device-centric framework, each using a GPU-resident (not the standard CPU-based) implementation of the respective search algorithm: \textbf{MATCHA w/ GPU MCTS}, \textbf{w/ GPU Gumbel MCTS}, and \textbf{w/ GPU SPO} \citep{macfarlane2024sposequentialmontecarlo}.
As \cref{fig:ablation_search_time} shows, MATCHA substantially improves wall-clock efficiency over the search baselines while maintaining sample efficiency comparable to GPU Gumbel MCTS.

To isolate the search procedure's efficiency and verify our time-complexity analysis in \cref{subsubsec:theory}, we compare the runtime scalability of the search algorithms in \cref{fig:search-scalability-k10}. 
We fix a trained SMAX \textit{3s\_vs\_5z} model with win rate near $1$ and measure the time each algorithm needs for one full search under different simulation budgets. 
Consistent with our analysis, the MCTS variants scale quadratically with the number of simulations, whereas PGFS is quasi-linear.
At $N=350$, PGFS is $15$--$20\times$ faster than the MCTS baselines.

\paragraph{Per-Component Profiling.} To understand the specific roles of our two contributions, the device-centric architecture and PGFS, we profile MATCHA, MATCHA w/ GPU Gumbel MCTS, and the CPU-centric baseline MAZero per component (see \cref{tab:profiling}).
For MAZero, the most time-consuming parts are the search-related components: data collection and reanalysis.
For MATCHA w/ GPU Gumbel MCTS, the device-centric architecture accelerates each component.
Reanalysis, which involves large-batch search, becomes the dominant bottleneck.
For full MATCHA, PGFS further reduces the cost of reanalysis and data collection, making them cheaper than optimization.
The dominant cost thus shifts from reanalysis to optimization, indicating that search is no longer the primary bottleneck.

\paragraph{Hyper-parameter Ablation.}
We also conducted hyperparameter ablations, including one on $\alpha$, which balances value score and prior constraint in PGFS (\cref{eq:action_selection}).
Results in \cref{appsubsec:alpha-ablation} show that PGFS is insensitive to $\alpha$: varying it from $50$ to $400$ yields nearly identical sample efficiency.

\section{Conclusions}
\label{sec:conclusions}
To bridge the gap between sample and wall-clock efficiency in MBMARL, we propose MATCHA, combining a system-level \textbf{device-centric end-to-end vectorized architecture} with an algorithm-level \textbf{Path-Guided Frontier Search}. Across SMAX, Overcooked, and MPE, it matches or exceeds strong MBMARL baselines in sample efficiency, with up to $16.3\times$ speedup on minimal GPU resources.

\paragraph{Limitations}
MATCHA targets the search and reanalysis bottlenecks in MBMARL. 
Once these costs are reduced, model optimization becomes the dominant bottleneck, limiting further speedup. 
Accelerating this less parallelizable stage remains an open challenge for future MBMARL systems.

\subsection*{AI use statement}
In this work, we used generative AI tools, including GPT and Claude, as research and writing assistants.
Specifically, we used these tools to draft, revise, polish, and translate portions of the manuscript; assist with code writing and debugging; check proofs and identify potential issues in mathematical derivations; create or refine figures; and help locate relevant literature.

The authors independently reviewed and revised all AI-assisted text, translations, and citations.
All AI-assisted code was manually inspected, tested, and executed by the authors.
Mathematical proofs and derivations were independently verified by the authors; AI tools were used only as an auxiliary aid for checking potential issues, not as a substitute for author verification.
The authors also verified the original sources and bibliographic information for all cited literature, and checked the data sources, accuracy, and final presentation of all figures.

We did not use generative AI tools to generate synthetic datasets, develop the core theoretical or conceptual framework, formulate the central mathematical claims, generate the final proofs, propose or refine the main hypotheses, design the research methodology or experiments, clean or reformat datasets, conduct qualitative or thematic data analysis, or interpret the main experimental results.
We take responsibility for the final content of this work, including all text, claims, code, figures, and other artifacts produced with the aid of generative AI.

\bibliography{main_ref}

@inproceedings{
liu2024efficient,
title={Efficient Multi-agent Reinforcement Learning by Planning},
author={Qihan Liu and Jianing Ye and Xiaoteng Ma and Jun Yang and Bin Liang and Chongjie Zhang},
booktitle={The Twelfth International Conference on Learning Representations},
year={2024},
url={https://openreview.net/forum?id=CpnKq3UJwp}
}

@inproceedings{
mei2023speedyzero,
title={SpeedyZero: Mastering Atari with Limited Data and Time},
author={Yixuan Mei and Jiaxuan Gao and Weirui Ye and Shaohuai Liu and Yang Gao and Yi Wu},
booktitle={The Eleventh International Conference on Learning Representations },
year={2023},
url={https://openreview.net/forum?id=Mg5CLXZgvLJ}
}

@misc{egorov2022scalablemultiagentmodelbasedreinforcement,
      title={Scalable Multi-Agent Model-Based Reinforcement Learning}, 
      author={Vladimir Egorov and Aleksei Shpilman},
      year={2022},
      eprint={2205.15023},
      archivePrefix={arXiv},
      primaryClass={cs.LG},
      url={https://arxiv.org/abs/2205.15023}, 
}

@article{
zhang2025decentralized,
title={Decentralized Transformers with Centralized Aggregation are Sample-Efficient Multi-Agent World Models},
author={Yang Zhang and Chenjia Bai and Bin Zhao and Junchi Yan and Xiu Li and Xuelong Li},
journal={Transactions on Machine Learning Research},
issn={2835-8856},
year={2025},
url={https://openreview.net/forum?id=xT8BEgXmVc},
note={}
}

@misc{zhang2025revisitingmultiagentworldmodeling,
      title={Revisiting Multi-Agent World Modeling from a Diffusion-Inspired Perspective}, 
      author={Yang Zhang and Xinran Li and Jianing Ye and Shuang Qiu and Delin Qu and Xiu Li and Chongjie Zhang and Chenjia Bai},
      year={2025},
      eprint={2505.20922},
      archivePrefix={arXiv},
      primaryClass={cs.MA},
      url={https://arxiv.org/abs/2505.20922}, 
}

@misc{wang2021qplexduplexduelingmultiagent,
      title={QPLEX: Duplex Dueling Multi-Agent Q-Learning}, 
      author={Jianhao Wang and Zhizhou Ren and Terry Liu and Yang Yu and Chongjie Zhang},
      year={2021},
      eprint={2008.01062},
      archivePrefix={arXiv},
      primaryClass={cs.LG},
      url={https://arxiv.org/abs/2008.01062}, 
}

@misc{rashid2018qmixmonotonicvaluefunction,
      title={QMIX: Monotonic Value Function Factorisation for Deep Multi-Agent Reinforcement Learning}, 
      author={Tabish Rashid and Mikayel Samvelyan and Christian Schroeder de Witt and Gregory Farquhar and Jakob Foerster and Shimon Whiteson},
      year={2018},
      eprint={1803.11485},
      archivePrefix={arXiv},
      primaryClass={cs.LG},
      url={https://arxiv.org/abs/1803.11485}, 
}

@misc{sunehag2017valuedecompositionnetworkscooperativemultiagent,
      title={Value-Decomposition Networks For Cooperative Multi-Agent Learning}, 
      author={Peter Sunehag and Guy Lever and Audrunas Gruslys and Wojciech Marian Czarnecki and Vinicius Zambaldi and Max Jaderberg and Marc Lanctot and Nicolas Sonnerat and Joel Z. Leibo and Karl Tuyls and Thore Graepel},
      year={2017},
      eprint={1706.05296},
      archivePrefix={arXiv},
      primaryClass={cs.AI},
      url={https://arxiv.org/abs/1706.05296}, 
}

@misc{yu2022surprisingeffectivenessppocooperative,
      title={The Surprising Effectiveness of PPO in Cooperative, Multi-Agent Games}, 
      author={Chao Yu and Akash Velu and Eugene Vinitsky and Jiaxuan Gao and Yu Wang and Alexandre Bayen and Yi Wu},
      year={2022},
      eprint={2103.01955},
      archivePrefix={arXiv},
      primaryClass={cs.LG},
      url={https://arxiv.org/abs/2103.01955}, 
}

@misc{ye2025globaloptimalitycooperativemarl,
      title={Towards Global Optimality in Cooperative MARL with the Transformation And Distillation Framework}, 
      author={Jianing Ye and Chenghao Li and Yongqiang Dou and Jianhao Wang and Guangwen Yang and Chongjie Zhang},
      year={2025},
      eprint={2207.11143},
      archivePrefix={arXiv},
      primaryClass={cs.MA},
      url={https://arxiv.org/abs/2207.11143}, 
}

@misc{ye2021masteringatarigameslimited,
      title={Mastering Atari Games with Limited Data}, 
      author={Weirui Ye and Shaohuai Liu and Thanard Kurutach and Pieter Abbeel and Yang Gao},
      year={2021},
      eprint={2111.00210},
      archivePrefix={arXiv},
      primaryClass={cs.LG},
      url={https://arxiv.org/abs/2111.00210}, 
}

@misc{wang2024efficientzerov2masteringdiscrete,
      title={EfficientZero V2: Mastering Discrete and Continuous Control with Limited Data}, 
      author={Shengjie Wang and Shaohuai Liu and Weirui Ye and Jiacheng You and Yang Gao},
      year={2024},
      eprint={2403.00564},
      archivePrefix={arXiv},
      primaryClass={cs.LG},
      url={https://arxiv.org/abs/2403.00564}, 
}

@misc{hafner2020dreamcontrollearningbehaviors,
      title={Dream to Control: Learning Behaviors by Latent Imagination}, 
      author={Danijar Hafner and Timothy Lillicrap and Jimmy Ba and Mohammad Norouzi},
      year={2020},
      eprint={1912.01603},
      archivePrefix={arXiv},
      primaryClass={cs.LG},
      url={https://arxiv.org/abs/1912.01603}, 
}

@article{Schrittwieser_2020,
   title={Mastering Atari, Go, chess and shogi by planning with a learned model},
   volume={588},
   ISSN={1476-4687},
   url={http://dx.doi.org/10.1038/s41586-020-03051-4},
   DOI={10.1038/s41586-020-03051-4},
   number={7839},
   journal={Nature},
   publisher={Springer Science and Business Media LLC},
   author={Schrittwieser, Julian and Antonoglou, Ioannis and Hubert, Thomas and Simonyan, Karen and Sifre, Laurent and Schmitt, Simon and Guez, Arthur and Lockhart, Edward and Hassabis, Demis and Graepel, Thore and Lillicrap, Timothy and Silver, David},
   year={2020},
   month=dec, pages={604–609} }

@inproceedings{
    flair2024jaxmarl,
    title={JaxMARL: Multi-Agent RL Environments and Algorithms in JAX},
    author={Alexander Rutherford and Benjamin Ellis and Matteo Gallici and Jonathan Cook and Andrei Lupu and Gar{\dh}ar Ingvarsson and Timon Willi and Ravi Hammond and Akbir Khan and Christian Schroeder de Witt and Alexandra Souly and Saptarashmi Bandyopadhyay and Mikayel Samvelyan and Minqi Jiang and Robert Tjarko Lange and Shimon Whiteson and Bruno Lacerda and Nick Hawes and Tim Rockt{\"a}schel and Chris Lu and Jakob Nicolaus Foerster},
    booktitle={The Thirty-eight Conference on Neural Information Processing Systems Datasets and Benchmarks Track},
    year={2024},
}

@misc{mahjoub2025sableperformantefficientscalable,
      title={Sable: a Performant, Efficient and Scalable Sequence Model for MARL}, 
      author={Omayma Mahjoub and Sasha Abramowitz and Ruan de Kock and Wiem Khlifi and Simon du Toit and Jemma Daniel and Louay Ben Nessir and Louise Beyers and Claude Formanek and Liam Clark and Arnu Pretorius},
      year={2025},
      eprint={2410.01706},
      archivePrefix={arXiv},
      primaryClass={cs.LG},
      url={https://arxiv.org/abs/2410.01706}, 
}

@misc{bonnet2024jumanji,
    title={Jumanji: a Diverse Suite of Scalable Reinforcement Learning Environments in JAX},
    author={Clément Bonnet and Daniel Luo and Donal Byrne and Shikha Surana and Sasha Abramowitz and Paul Duckworth and Vincent Coyette and Laurence I. Midgley and Elshadai Tegegn and Tristan Kalloniatis and Omayma Mahjoub and Matthew Macfarlane and Andries P. Smit and Nathan Grinsztajn and Raphael Boige and Cemlyn N. Waters and Mohamed A. Mimouni and Ulrich A. Mbou Sob and Ruan de Kock and Siddarth Singh and Daniel Furelos-Blanco and Victor Le and Arnu Pretorius and Alexandre Laterre},
    year={2024},
    eprint={2306.09884},
    url={https://arxiv.org/abs/2306.09884},
    archivePrefix={arXiv},
    primaryClass={cs.LG}
}

@misc{nair2015massivelyparallelmethodsdeep,
      title={Massively Parallel Methods for Deep Reinforcement Learning}, 
      author={Arun Nair and Praveen Srinivasan and Sam Blackwell and Cagdas Alcicek and Rory Fearon and Alessandro De Maria and Vedavyas Panneershelvam and Mustafa Suleyman and Charles Beattie and Stig Petersen and Shane Legg and Volodymyr Mnih and Koray Kavukcuoglu and David Silver},
      year={2015},
      eprint={1507.04296},
      archivePrefix={arXiv},
      primaryClass={cs.LG},
      url={https://arxiv.org/abs/1507.04296}, 
}

@misc{espeholt2018impalascalabledistributeddeeprl,
      title={IMPALA: Scalable Distributed Deep-RL with Importance Weighted Actor-Learner Architectures}, 
      author={Lasse Espeholt and Hubert Soyer and Remi Munos and Karen Simonyan and Volodymir Mnih and Tom Ward and Yotam Doron and Vlad Firoiu and Tim Harley and Iain Dunning and Shane Legg and Koray Kavukcuoglu},
      year={2018},
      eprint={1802.01561},
      archivePrefix={arXiv},
      primaryClass={cs.LG},
      url={https://arxiv.org/abs/1802.01561}, 
}

@misc{horgan2018distributedprioritizedexperiencereplay,
      title={Distributed Prioritized Experience Replay}, 
      author={Dan Horgan and John Quan and David Budden and Gabriel Barth-Maron and Matteo Hessel and Hado van Hasselt and David Silver},
      year={2018},
      eprint={1803.00933},
      archivePrefix={arXiv},
      primaryClass={cs.LG},
      url={https://arxiv.org/abs/1803.00933}, 
}

@misc{ye2022spendingthinkingtimewisely,
      title={Spending Thinking Time Wisely: Accelerating MCTS with Virtual Expansions}, 
      author={Weirui Ye and Pieter Abbeel and Yang Gao},
      year={2022},
      eprint={2210.12628},
      archivePrefix={arXiv},
      primaryClass={cs.AI},
      url={https://arxiv.org/abs/2210.12628}, 
}

@inproceedings{
danihelka2022policy,
title={Policy improvement by planning with Gumbel},
author={Ivo Danihelka and Arthur Guez and Julian Schrittwieser and David Silver},
booktitle={International Conference on Learning Representations},
year={2022},
url={https://openreview.net/forum?id=bERaNdoegnO}
}

@misc{hubert2021learningplanningcomplexaction,
      title={Learning and Planning in Complex Action Spaces}, 
      author={Thomas Hubert and Julian Schrittwieser and Ioannis Antonoglou and Mohammadamin Barekatain and Simon Schmitt and David Silver},
      year={2021},
      eprint={2104.06303},
      archivePrefix={arXiv},
      primaryClass={cs.LG},
      url={https://arxiv.org/abs/2104.06303}, 
}

@misc{brax2021github,
  author = {C. Daniel Freeman and Erik Frey and Anton Raichuk and Sertan Girgin and Igor Mordatch and Olivier Bachem},
  title = {Brax - A Differentiable Physics Engine for Large Scale Rigid Body Simulation},
  howpublished = {\url{http://github.com/google/brax}},
  note = {Version 0.14.0},
  year = {2021},
}

@misc{jaxrl,
  author = {Kostrikov, Ilya},
  doi = {10.5281/zenodo.5535154},
  month = {10},
  title = {{JAXRL: Implementations of Reinforcement Learning algorithms in JAX}},
  url = {https://github.com/ikostrikov/jaxrl},
  year = {2021}
}

@article{samvelyan2019smac,
  title={The starcraft multi-agent challenge},
  author={Samvelyan, Mikayel and Rashid, Tabish and De Witt, Christian Schroeder and Farquhar, Gregory and Nardelli, Nantas and Rudner, Tim GJ and Hung, Chia-Man and Torr, Philip HS and Foerster, Jakob and Whiteson, Shimon},
  journal={arXiv preprint arXiv:1902.04043},
  year={2019}
}

@article{carroll2019overcooked,
  title={On the utility of learning about humans for human-ai coordination},
  author={Carroll, Micah and Shah, Rohin and Ho, Mark K and Griffiths, Tom and Seshia, Sanjit and Abbeel, Pieter and Dragan, Anca},
  journal={Advances in neural information processing systems},
  volume={32},
  year={2019}
}

@article{lowe2017mpe,
  title={Multi-agent actor-critic for mixed cooperative-competitive environments},
  author={Lowe, Ryan and Wu, Yi and Tamar, Aviv and Harb, Jean and Pieter Abbeel, OpenAI and Mordatch, Igor},
  journal={Advances in neural information processing systems},
  volume={30},
  year={2017}
}

@misc{hessel2021podracerarchitecturesscalablereinforcement,
      title={Podracer architectures for scalable Reinforcement Learning}, 
      author={Matteo Hessel and Manuel Kroiss and Aidan Clark and Iurii Kemaev and John Quan and Thomas Keck and Fabio Viola and Hado van Hasselt},
      year={2021},
      eprint={2104.06272},
      archivePrefix={arXiv},
      primaryClass={cs.LG},
      url={https://arxiv.org/abs/2104.06272}, 
}

@misc{toledo2024stoix,
    title={Stoix: Distributed Single-Agent Reinforcement Learning End-to-End in JAX},
    doi = {10.5281/zenodo.10916257},
    author={Edan Toledo},
    month = apr,
    year = {2024},
    url = {https://github.com/EdanToledo/Stoix},
}

@misc{gymnax2022github,
  author = {Robert Tjarko Lange},
  title = {{gymnax}: A {JAX}-based Reinforcement Learning Environment Library},
  url = {http://github.com/RobertTLange/gymnax},
  version = {0.0.4},
  year = {2022},
}

@misc{dekock2023mavaresearchlibrarydistributed,
      title={Mava: a research library for distributed multi-agent reinforcement learning in JAX}, 
      author={Ruan de Kock and Omayma Mahjoub and Sasha Abramowitz and Wiem Khlifi and Callum Rhys Tilbury and Claude Formanek and Andries Smit and Arnu Pretorius},
      year={2023},
      eprint={2107.01460},
      archivePrefix={arXiv},
      primaryClass={cs.LG},
      url={https://arxiv.org/abs/2107.01460}, 
}

@inproceedings{koyamada2023pgx,
  title={Pgx: Hardware-Accelerated Parallel Game Simulators for Reinforcement Learning},
  author={Koyamada, Sotetsu and Okano, Shinri and Nishimori, Soichiro and Murata, Yu and Habara, Keigo and Kita, Haruka and Ishii, Shin},
  booktitle={Advances in Neural Information Processing Systems},
  pages={45716--45743},
  volume={36},
  year={2023}
}

@inproceedings{
radji2026octax,
title={Octax: Accelerated {CHIP}-8 Arcade Environments for Reinforcement Learning in {JAX}},
author={Waris Radji and Thomas Michel and Hector Piteau},
booktitle={The Fourteenth International Conference on Learning Representations},
year={2026},
url={https://openreview.net/forum?id=BdeUsYlyIf}
}

@article{DBLP:journals/corr/abs-1712-08650,
  author       = {Pierre H. Richemond and
                  Brendan Maginnis},
  title        = {A short variational proof of equivalence between policy gradients
                  and soft {Q} learning},
  journal      = {CoRR},
  volume       = {abs/1712.08650},
  year         = {2017},
  url          = {http://arxiv.org/abs/1712.08650},
  eprinttype   = {arXiv},
  eprint       = {1712.08650},
  bibsource    = {dblp computer science bibliography, https://dblp.org}
}

@article{Fenwick1994AND,
  title={A new data structure for cumulative frequency tables},
  author={Peter M. Fenwick},
  journal={Software: Practice and Experience},
  year={1994},
  volume={24},
  url={https://api.semanticscholar.org/CorpusID:7519761}
}

@inproceedings{
doro2023sampleefficient,
title={Sample-Efficient Reinforcement Learning by Breaking the Replay Ratio Barrier},
author={Pierluca D'Oro and Max Schwarzer and Evgenii Nikishin and Pierre-Luc Bacon and Marc G Bellemare and Aaron Courville},
booktitle={The Eleventh International Conference on Learning Representations },
year={2023},
url={https://openreview.net/forum?id=OpC-9aBBVJe}
}

@misc{macfarlane2024sposequentialmontecarlo,
      title={SPO: Sequential Monte Carlo Policy Optimisation}, 
      author={Matthew V Macfarlane and Edan Toledo and Donal Byrne and Paul Duckworth and Alexandre Laterre},
      year={2024},
      eprint={2402.07963},
      archivePrefix={arXiv},
      primaryClass={cs.AI},
      url={https://arxiv.org/abs/2402.07963}, 
}

@software{deepmind2020jax,
  title = {The {D}eep{M}ind {JAX} {E}cosystem},
  author = {DeepMind and Babuschkin, Igor and Baumli, Kate and Bell, Alison and Bhupatiraju, Surya and Bruce, Jake and Buchlovsky, Peter and Budden, David and Cai, Trevor and Clark, Aidan and Danihelka, Ivo and Dedieu, Antoine and Fantacci, Claudio and Godwin, Jonathan and Jones, Chris and Hemsley, Ross and Hennigan, Tom and Hessel, Matteo and Hou, Shaobo and Kapturowski, Steven and Keck, Thomas and Kemaev, Iurii and King, Michael and Kunesch, Markus and Martens, Lena and Merzic, Hamza and Mikulik, Vladimir and Norman, Tamara and Papamakarios, George and Quan, John and Ring, Roman and Ruiz, Francisco and Sanchez, Alvaro and Sartran, Laurent and Schneider, Rosalia and Sezener, Eren and Spencer, Stephen and Srinivasan, Srivatsan and Stanojevi\'{c}, Milo\v{s} and Stokowiec, Wojciech and Wang, Luyu and Zhou, Guangyao and Viola, Fabio},
  url = {http://github.com/deepmind},
  year = {2020},
}

@article{nguyen2020deep,
  title={Deep reinforcement learning for multiagent systems: A review of challenges, solutions, and applications},
  author={Nguyen, Thanh Thi and Nguyen, Ngoc Duy and Nahavandi, Saeid},
  journal={IEEE transactions on cybernetics},
  volume={50},
  number={9},
  pages={3826--3839},
  year={2020},
  publisher={IEEE}
}

@inproceedings{wu2023modelsasagents,
  title={Models as agents: Optimizing multi-step predictions of interactive local models in model-based multi-agent reinforcement learning},
  author={Wu, Zifan and Yu, Chao and Chen, Chen and Hao, Jianye and Zhuo, Hankz Hankui},
  booktitle={Proceedings of the AAAI Conference on Artificial Intelligence},
  volume={37},
  pages={10435--10443},
  year={2023}
}

@inproceedings{hernandez2020very,
  title={A very condensed survey and critique of multiagent deep reinforcement learning},
  author={Hernandez-Leal, Pablo and Kartal, Bilal and Taylor, Matthew E},
  booktitle={Proceedings of the 19th international conference on autonomous agents and multiagent systems},
  pages={2146--2148},
  year={2020}
}

@inproceedings{ha18worldmodels,
 author = {Ha, David and Schmidhuber, J\"{u}rgen},
 booktitle = {Advances in Neural Information Processing Systems},
 editor = {S. Bengio and H. Wallach and H. Larochelle and K. Grauman and N. Cesa-Bianchi and R. Garnett},
 pages = {},
 publisher = {Curran Associates, Inc.},
 title = {Recurrent World Models Facilitate Policy Evolution},
 url = {https://proceedings.neurips.cc/paper_files/paper/2018/file/2de5d16682c3c35007e4e92982f1a2ba-Paper.pdf},
 volume = {31},
 year = {2018}
}

@article{hafner2020dreamerv2,
  title={Mastering Atari with Discrete World Models},
  author={Hafner, Danijar and Lillicrap, Timothy and Norouzi, Mohammad and Ba, Jimmy},
  journal={arXiv preprint arXiv:2010.02193},
  year={2020}
}

@article{hafner2025dreamerv3,
  title={Mastering diverse control tasks through world models},
  author={Hafner, Danijar and Pasukonis, Jurgis and Ba, Jimmy and Lillicrap, Timothy},
  journal={Nature},
  pages={1--7},
  year={2025},
  publisher={Nature Publishing Group}
}

@InProceedings{son19qtran,
  title = 	 {{QTRAN}: Learning to Factorize with Transformation for Cooperative Multi-Agent Reinforcement Learning},
  author =       {Son, Kyunghwan and Kim, Daewoo and Kang, Wan Ju and Hostallero, David Earl and Yi, Yung},
  booktitle = 	 {Proceedings of the 36th International Conference on Machine Learning},
  year = 	 {2019},
  series = 	 {Proceedings of Machine Learning Research},
  publisher =    {PMLR},
  url = 	 {https://proceedings.mlr.press/v97/son19a.html}
}

@inproceedings{Foerster18coma,
    author = {Foerster, Jakob N. and Farquhar, Gregory and Afouras, Triantafyllos and Nardelli, Nantas and Whiteson, Shimon},
    title = {Counterfactual multi-agent policy gradients},
    year = {2018},
    booktitle = {Proceedings of the Thirty-Second AAAI Conference on Artificial Intelligence},
}

@InProceedings{iqbal19maac,
  title =    {Actor-Attention-Critic for Multi-Agent Reinforcement Learning},
  author =   {Iqbal, Shariq and Sha, Fei},
  booktitle =    {Proceedings of the 36th International Conference on Machine Learning},
  year =     {2019},
  series =   {Proceedings of Machine Learning Research},
  publisher =    {PMLR},
  url =      {http://proceedings.mlr.press/v97/iqbal19a.html},
}

@inproceedings{Ryu2019MultiAgentAW,
  title={Multi-agent actor-critic with hierarchical graph attention network},
  author={Ryu, Heechang and Shin, Hayong and Park, Jinkyoo},
  booktitle={Proceedings of the AAAI Conference on Artificial Intelligence},
  year={2020}
}

@inproceedings{liu20maga,
  author       = {Yong Liu and
                  Weixun Wang and
                  Yujing Hu and
                  Jianye Hao and
                  Xingguo Chen and
                  Yang Gao},
  title        = {Multi-Agent Game Abstraction via Graph Attention Neural Network},
  booktitle    = {The Thirty-Fourth {AAAI} Conference on Artificial Intelligence},
  year         = {2020},
  url          = {https://doi.org/10.1609/aaai.v34i05.6211},
  doi          = {10.1609/AAAI.V34I05.6211},
}

@inproceedings{kuba2021trust,
  title={Trust Region Policy Optimisation in Multi-Agent Reinforcement Learning},
  author={Kuba, Jakub Grudzien and Chen, Ruiqing and Wen, Muning and Wen, Ying and Sun, Fanglei and Wang, Jun and Yang, Yaodong},
  booktitle={International Conference on Learning Representations},
  year={2021}
}

@inproceedings{peng2021facmac,
    title={{FACMAC}: Factored Multi-Agent Centralised Policy Gradients},
    author={Bei Peng and Tabish Rashid and Christian Schroeder de Witt and Pierre-Alexandre Kamienny and Philip Torr and Wendelin Boehmer and Shimon Whiteson},
    booktitle={Advances in Neural Information Processing Systems},
    year={2021},
    url={https://openreview.net/forum?id=WxH774N0mEu}
}

@article{zhang2024read,
  title={Towards Efficient LLM Grounding for Embodied Multi-Agent Collaboration},
  author={Zhang, Yang and Yang, Shixin and Bai, Chenjia and Wu, Fei and Li, Xiu and Li, Xuelong and Wang, Zhen},
  journal={arXiv preprint arXiv:2405.14314},
  year={2024}
}
\bibliographystyle{iclr2027_conference}

\clearpage
\appendix

\section{Additional Experimental Results}

\subsection{Performance on Multi-Particle Environment (MPE)}
\label{app:mpe_results}

We further evaluate MATCHA on the Multi-Particle Environment (MPE) to assess its capabilities in continuous observation spaces and communication-oriented tasks.

\begin{figure}[htb]
    \centering
    \includegraphics[width=\textwidth]{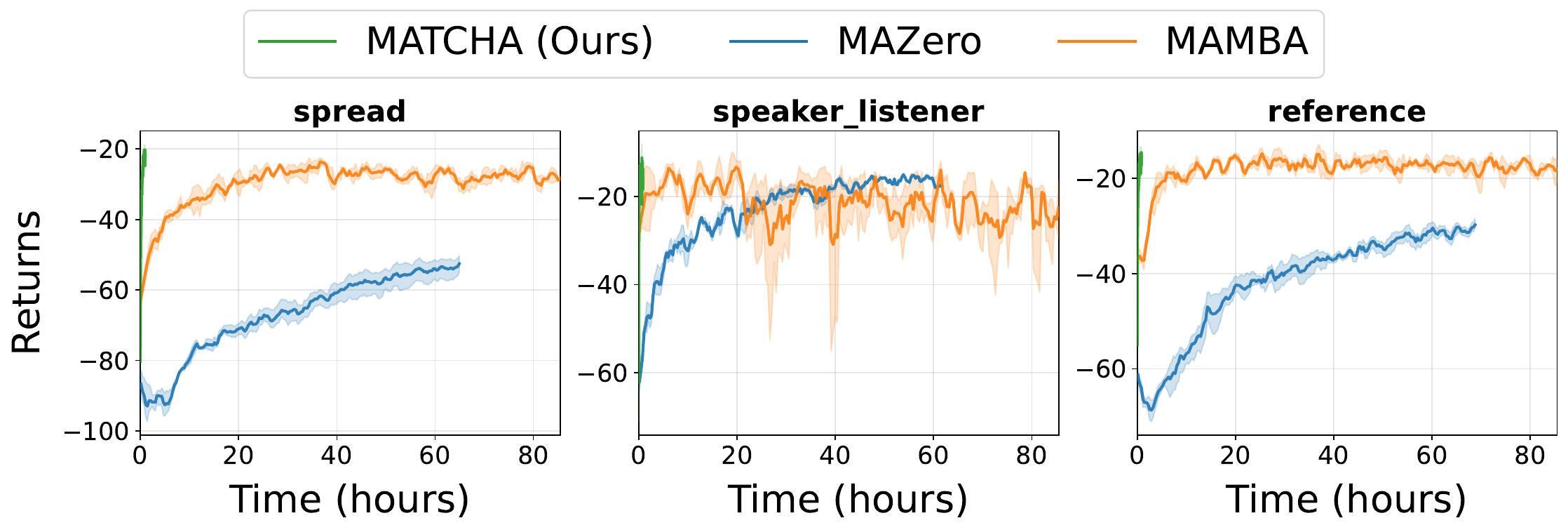}
    \caption{\textbf{Wall-clock Efficiency on MPE.} Comparison of MATCHA against model-based baselines on MPE scenarios.}
    \label{fig:mpe_results_time_efficiency_full}
\end{figure}

In terms of wall-clock efficiency, \cref{fig:mpe_results_time_efficiency_full} shows that MATCHA continues to dominate both model-based baselines, MAZero and MAMBA, on MPE.

\begin{figure}[htb]
    \centering
    \includegraphics[width=\textwidth]{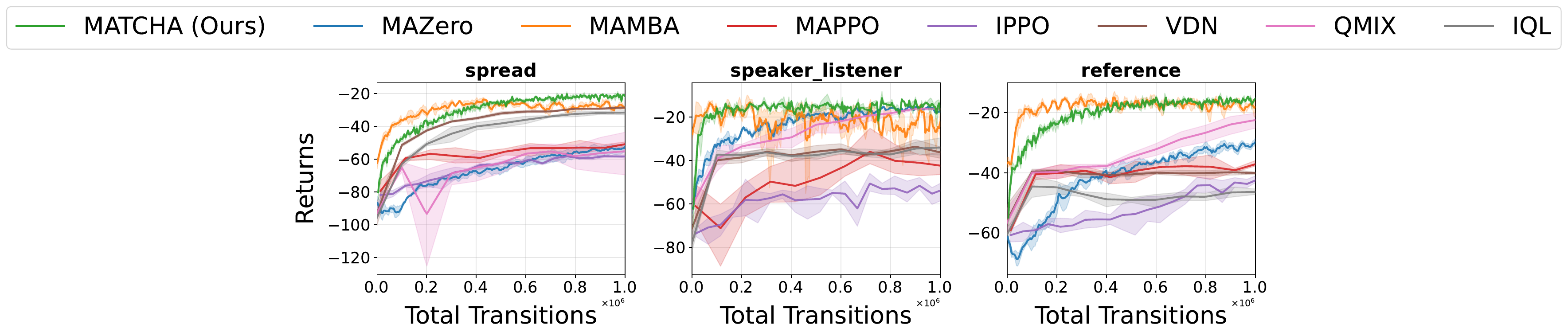}
    \caption{\textbf{Sample Efficiency on MPE.} Comparison of MATCHA against model-based and model-free baselines on MPE scenarios.}
    \label{fig:mpe_results_sample_efficiency}
\end{figure}

As shown in \cref{fig:mpe_results_sample_efficiency}, MATCHA achieves substantially higher sample efficiency than all model-free baselines (MAPPO, IPPO, QMIX, VDN, IQL) as well as the model-based baseline MAZero, across all three tested scenarios.
MAMBA improves faster during early training, but consistently converges to a worse (or, at best, comparable) asymptotic performance than MATCHA.
\begin{itemize}
    \item \textbf{Spread:} Requires agents to cover landmarks while avoiding collisions. MATCHA converges faster and achieves a higher collective return, indicating effective spatial coordination.
    \item \textbf{Speaker-Listener:} A communication-heavy task where one agent must signal a target landmark to another. MATCHA's rapid convergence suggests that the search mechanism effectively facilitates the emergence of communication protocols.
    \item \textbf{Reference:} Agents must move to a target referenced by another agent. MATCHA shows significantly lower variance and higher asymptotic performance compared to QMIX and VDN.
\end{itemize}

\subsection{Full Wall-Clock Efficiency Performance Results on SMAX and Overcooked}
\label{app:wall-clock}

\begin{figure*}[h]
    \centering
    \includegraphics[width=\textwidth]{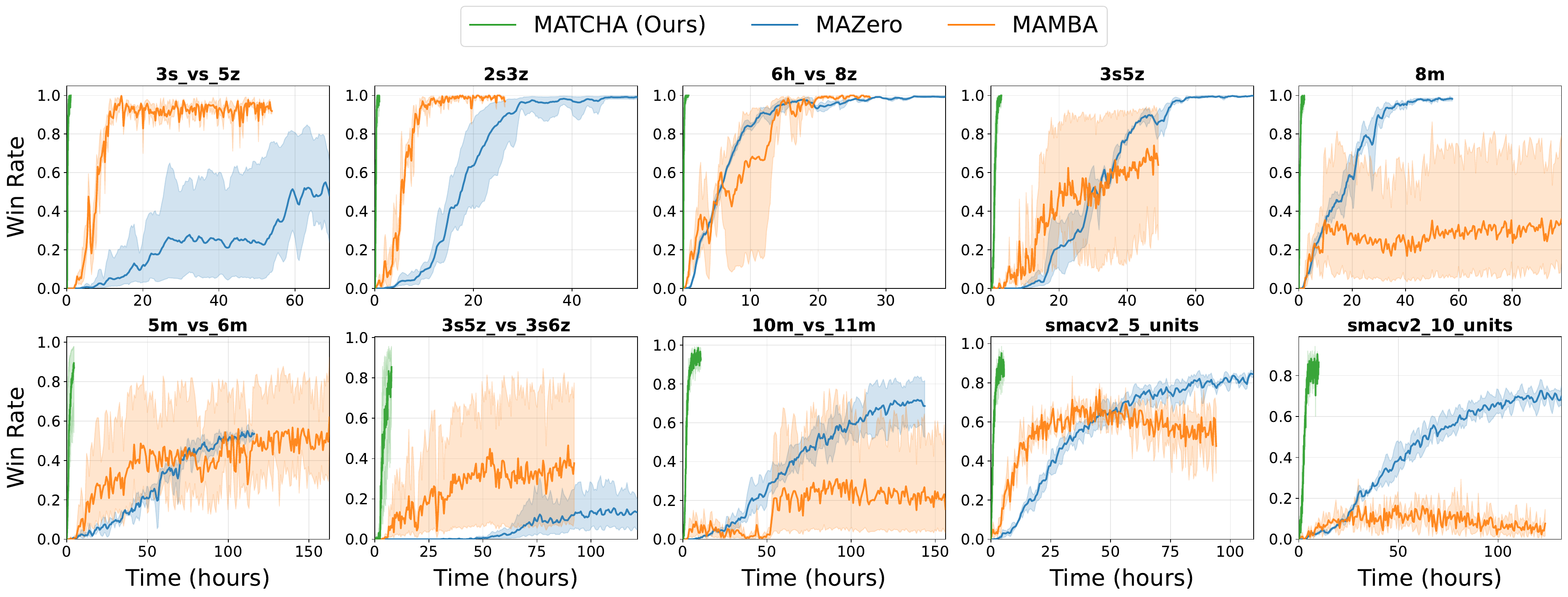}
    \caption{\textbf{Full Wall-clock efficiency performance (win rate vs. time) on SMAX.}}
    \label{fig:smax_full_time}
\end{figure*}

\begin{figure*}[h]
    \centering
    \includegraphics[width=\textwidth]{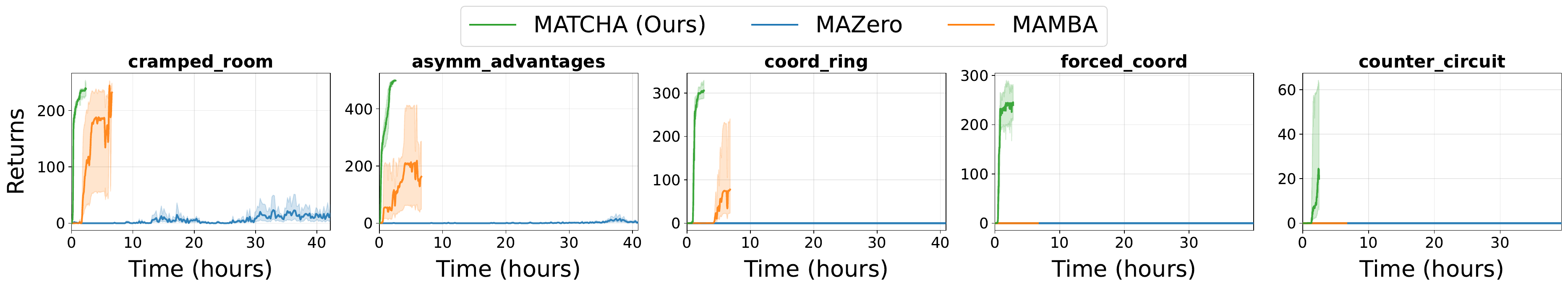}
    \caption{\textbf{Full Wall-clock efficiency performance (returns vs. time) on Overcooked.}}
    \label{fig:overcooked_full_time}
\end{figure*}

\paragraph{Resource settings.}
Unless otherwise noted (i.e., the GPU-scalability study in \cref{appsubsec:gpu-scalability} and the CPU-centric SMAC results in \cref{appsubsec:smac-cpu}, which use the separate hardware described in their respective sections), all experiments reported in the main paper use the same hardware configuration: $2\times$ NVIDIA RTX 5090 GPUs, a 48-core AMD Ryzen Threadripper PRO 5965WX CPU, and 126\,GB of RAM.
Each experiment uses a single GPU by default, including the wall-clock efficiency results on SMAX and Overcooked (\cref{fig:smax_time,fig:overcooked_time,fig:smax_full_time,fig:overcooked_full_time}); the sole exception is the per-component profiling result (\cref{tab:profiling}), which uses both GPUs.
We profile on two GPUs rather than one because cross-device aggregation, where a centralized learner aggregates reanalyzed batches produced by each device's on-device chain (\cref{subsec:arch}), is a defining feature of MATCHA's architecture; profiling on a single GPU would collapse to the degenerate case where the centralized learner has nothing to aggregate across, and would not reflect the per-component costs of this cross-device execution.
All compared methods within a given experiment are evaluated on the same hardware configuration to ensure a fair comparison.

\cref{fig:smax_time,fig:overcooked_time} present time-truncated results to facilitate visual comparison across methods, rather than letting slower approaches such as MAMBA and MAZero until reach the same number of environment steps as MATCHA. This section reports the full wall-clock results without any time truncation, as shown in \cref{fig:smax_full_time,fig:overcooked_full_time}.


\subsection{Compatibility with CPU-Centric Environments: Results on the Original SMAC}
\label{appsubsec:smac-cpu}

The experiments in the main paper are conducted on fully vectorized, GPU-native environments (SMAX, Overcooked, MPE). To verify that MATCHA's wall-clock efficiency advantage is not specific to such GPU-vectorized environments, we further evaluate it on the original Python/C++-backed SMAC \citep{samvelyan2019smac} environment, which is CPU-centric and does not support GPU-vectorized simulation. We test on the \textit{3s\_vs\_5z} scenario and compare MATCHA against the model-based baselines MAZero and MAMBA.

\begin{figure}[h]
    \centering
    \includegraphics[width=0.6\linewidth]{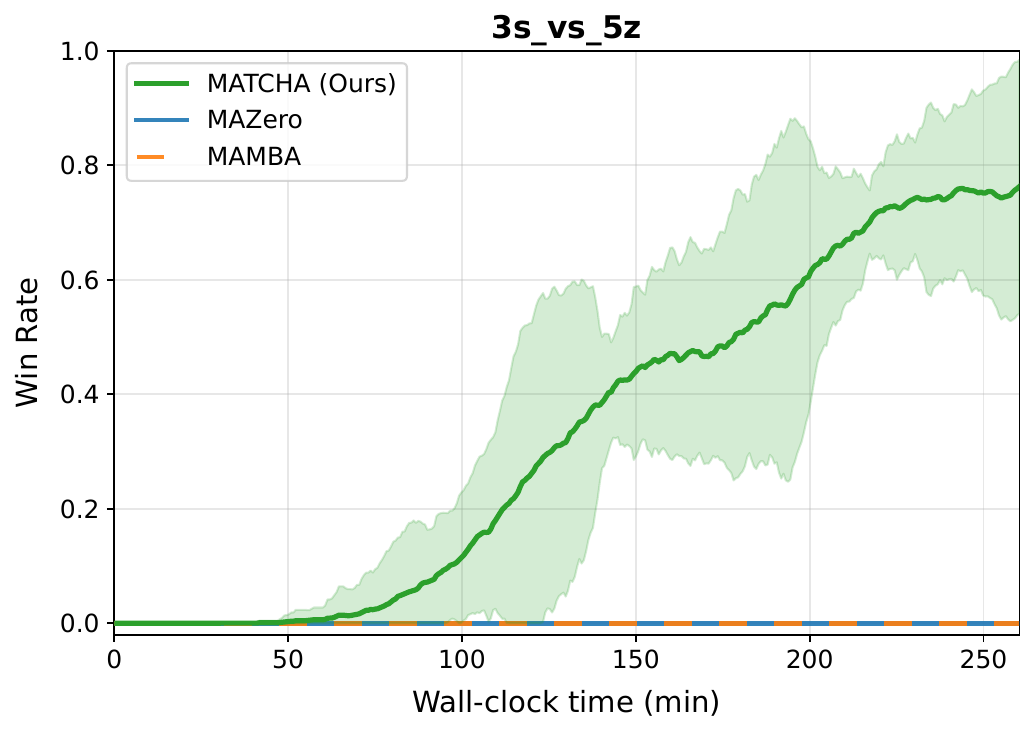}
    \caption{\textbf{Wall-Clock Efficiency on the Original SMAC \textit{3s\_vs\_5z}.} Win rate vs. wall-clock time on the CPU-centric, non-vectorized SMAC environment.}
    \label{fig:smac-speed}
\end{figure}

As shown in \cref{fig:smac-speed}, within 250 minutes MATCHA already reaches a win rate close to $80\%$, whereas both MAZero and MAMBA remain at a $0\%$ win rate throughout the entire run; these baselines typically require a week or longer to reach a comparable win rate on this task. This confirms that MATCHA's device-centric design yields substantial wall-clock speedups even when interfacing with a legacy, CPU-centric environment, rather than relying solely on the environment itself being GPU-vectorized.

\subsection{Ablation Study for Update-to-Data Ratio (UTD)}
\label{appsubsec:UTD}

In this section, we study the effect of the update-to-data ratio (UTD) on both the sample efficiency and wall-clock efficiency of the algorithm.

UTD is an important factor that affects both the sample efficiency and the running time of an algorithm \citep{doro2023sampleefficient}.
It measures the average number of gradient updates performed per environment interaction. Its impact on algorithmic performance can be understood from two aspects.

First, within a certain range, increasing the UTD generally improves sample efficiency, since the algorithm performs more optimization steps on the same amount of collected data.
Second, increasing the UTD also increases the wall-clock training time.
Environment interactions are usually easy to parallelize, whereas model updates are inherently more sequential.
When the UTD becomes sufficiently large, model learning, including reanalysis and optimization in our algorithm, becomes the main bottleneck.
Consequently, the time required to process the same number of environment interactions grows approximately linearly with the UTD.

\begin{figure*}[h]
    \centering
    \includegraphics[width=\textwidth]{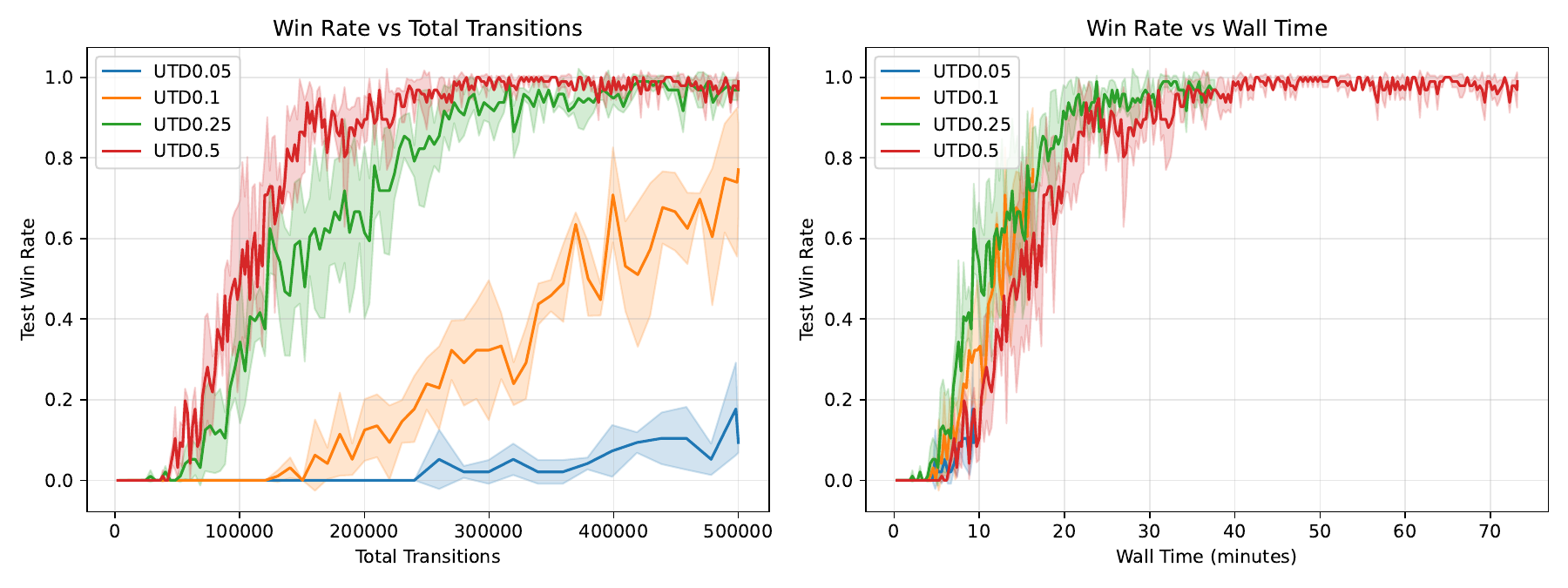}
    \caption{\textbf{Ablation Study for Different UTD on SMAX \textsc{3s\_vs\_5z}.}}
    \label{fig:utd_ablation}
\end{figure*}

We conduct ablation experiments with different UTD values, as shown in \cref{fig:utd_ablation}.
The results confirm that, within a reasonable range, a higher UTD improves sample efficiency but also increases the total training time.
However, a longer total training time does not necessarily imply worse wall-clock efficiency.
As shown in the figure, different UTD values achieve comparable wall-clock efficiency, and lower UTD values are even slightly more efficient.
This is because, for the same number of model updates, a lower UTD allows the algorithm to incorporate fresher data from the environment.
Since the running time is approximately proportional to the number of model updates, lower UTD values can lead to slightly better wall-clock efficiency.

\subsection{Comparison with DDP-style Scaling (e.g., Anakin)}
\label{appsubsec:DDP}

As discussed in \cref{subsec:arch}, a key difference between our framework and a DDP-style framework lies in how computation is scaled across devices.
A DDP-style framework scales out the entire learning process: multiple workers collect data and perform gradient updates in parallel, after which their gradients or model parameters are averaged.
Although this design increases the amount of data collected per synchronization step, it still produces only one effective optimizer update after aggregation.
Therefore, if the number of effective optimizer updates is kept fixed, increasing the number of workers reduces the effective UTD approximately in inverse proportion to the number of workers. This can substantially degrade sample efficiency.

In contrast, our framework only scales out the pipeline from data collection to reanalysis.
The model optimizer sequentially consumes the reanalyzed batches produced by different workers.
This design allows us to increase the throughput of data generation and reanalysis while keeping the UTD unchanged, thereby preserving sample efficiency.

In summary, a DDP-style framework improves the throughput of environment interactions by scaling out the entire learning process, but it reduces the effective UTD and therefore may fail to maintain sample efficiency.
By contrast, our framework scales out only the pipeline from data collection to reanalysis.
Together with a faster search algorithm, this effectively increases the amount of useful training data available within each optimization cycle, improves training throughput, and preserves sample efficiency.

\subsection{GPU Scalability of MBMARL Algorithms}
\label{appsubsec:gpu-scalability}

\begin{figure}[h]
    \centering
    \includegraphics[width=0.6\linewidth]{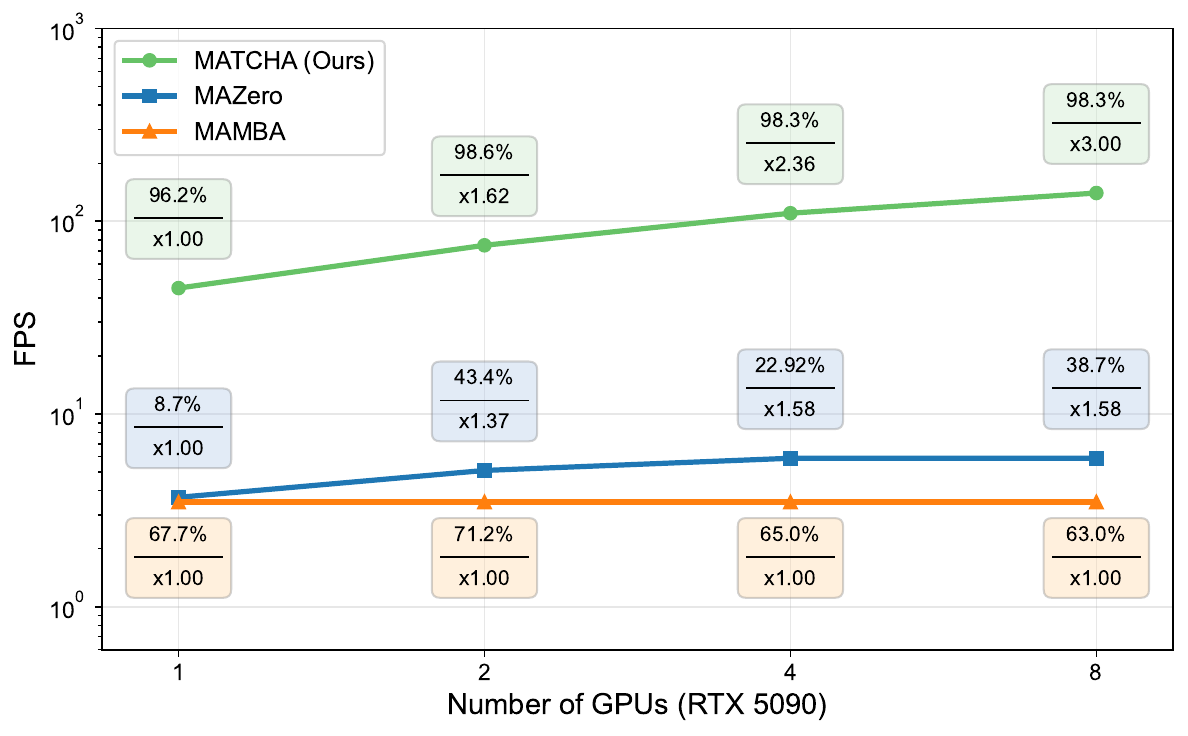}
    \caption{\textbf{Scalability of MBMARL algorithms.} FPS and truncated win rate on SMAX \textit{6h\_vs\_8z} across model-based MARL methods under different numbers of RTX 5090 GPUs; annotations show win rate at $80\mathrm{K}$ environment steps above the line and speedup over the single-GPU setting below it.}
    \label{fig:mbmarl-gpu-scalability}
\end{figure}

Another key property of an efficient learning system is that it can leverage additional compute to run faster \emph{without} sacrificing performance.
To evaluate GPU scalability, we scale MATCHA and model-based baselines from 1 to 8 GPUs on a single server equipped with $8\times$ NVIDIA RTX 5090 GPUs and 256\,GB of total VRAM (32\,GB per card), paired with $4\times$ AMD EPYC 9654 96-core processors (384 cores total) and 1032\,GB of RAM.
As shown in \cref{fig:mbmarl-gpu-scalability}, MATCHA achieves higher throughput while preserving sample efficiency as GPUs are added, demonstrating strong scalability.
In contrast, MAMBA does not meaningfully benefit from GPU scaling: its highly sequential control flow leaves additional GPUs largely idle.
MAZero can take advantage of extra RAM/VRAM by increasing the number of reanalysis workers, leading to modest scaling from 1 to 4 GPUs.
However, using more reanalysis workers causes CPU thread resources to become saturated, preventing MAZero from scaling effectively in an 8-GPU environment.

\subsection{Comparison between CPU buffer and GPU buffer}

Our framework MATCHA supports the use of both CPU and GPU buffers.
The primary reason we choose a GPU buffer as default is that it completely avoids CPU–GPU communication, allowing the entire algorithm to be compiled into a single kernel and executed fully on the GPU, thereby best reflecting our device-centric design philosophy.
As a result, aside from the initial few minutes required for compilation with CPU involvement, the subsequent execution speed depends only on the GPU.
This enables full reproducibility across machines with the same GPU model, independent of differences in CPU or memory.

On the other hand, for tasks that require large buffer capacity, such as image-based tasks, we also support the use of a CPU buffer.
In fact, since the buffer implementation is based on a sum tree, a data structure that is more CPU-friendly.
Therefore the buffer execution on CPU can even be faster than on GPU (\cref{tab:buffer_time_comparison}).
However, this comes at the cost of additional CPU–GPU communication overhead, which may vary across different machines.
However, even for CPU buffer, the buffer-related time cost is not the bottleneck (see \cref{tab:profiling}), since the CPU--GPU communication induced by the replay buffer is sparse.

\begin{table}[h]
\centering
\caption{Per-step time cost comparison between CPU and GPU buffers.}
\label{tab:buffer_time_comparison}
\begin{tabular}{lccc}
\toprule
\multirow{2}{*}{Buffer type} 
& \multicolumn{3}{c}{Time per training step (sec)} \\
\cmidrule(lr){2-4}
& Buffer operations & CPU--GPU communication & Total \\
\midrule
CPU buffer & 0.000018025 & 0.00223 & 0.002248 \\
GPU buffer & 0.0000515   & 0       & 0.0000515 \\
\bottomrule
\end{tabular}
\end{table}

\subsection{Sensitivity Test for \texorpdfstring{$\alpha$}{alpha}}
\label{appsubsec:alpha-ablation}

\begin{figure*}[h]
    \centering
    \includegraphics[width=\textwidth]{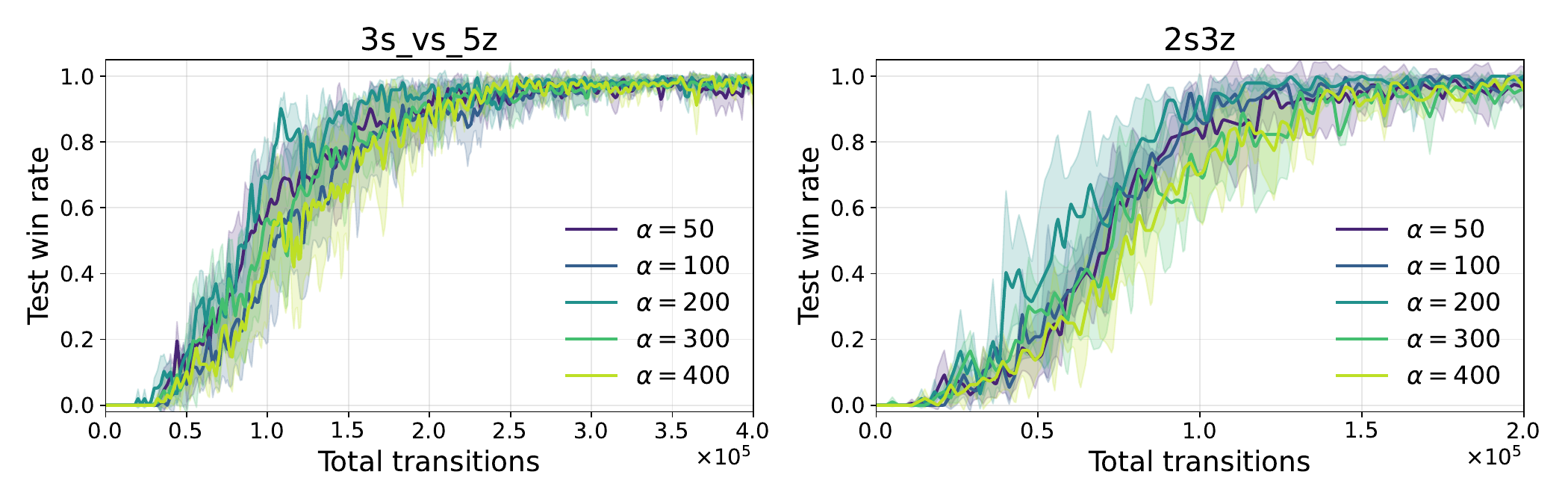}
    \caption{\textbf{Ablation Study for Different $\alpha$s on two SMAX Maps.}}
    \label{fig:alpha_ablation}
\end{figure*}

We further evaluate the sensitivity of PGFS to the hyperparameter $\alpha$, which controls the trade-off between the value score and the prior constraint. Specifically, we conduct experiments on two SMAX maps, \texttt{3s\_vs\_5z} and \texttt{2s3z}, with $\alpha \in \{50, 100, 200, 300, 400\}$. As shown in Figure~\ref{fig:alpha_ablation}, the learning curves under different choices of $\alpha$ exhibit similar sample efficiency and converge to comparable performance. This suggests that PGFS is not sensitive to the precise value of $\alpha$ within a broad range, demonstrating the robustness of the proposed method to this hyperparameter.

\clearpage
\section{Path-Guided Frontier Search}
\label{appsec:PGFS}

\subsection{Algorithm Description}

The Path-Guided Frontier Search (PGFS) algorithm is described in \cref{alg:PGFS_theoretical}.
The learned model is $\mathcal M_\theta=(v_\theta,T_\theta,\gamma_\theta)$, where $\gamma_\theta(x,a)\in[0,1]$ predicts the effective discount for action $a$ at $x$, including any continuation or termination factor.
For the resulting child $y$, write $\gamma(x,y)=\gamma_\theta(x,a)$.
The path statistic $d(x)$ is the product of these edge discounts along the root-to-$x$ path.

\begin{algorithm}[htb]
\caption{Path-Guided Frontier Search}
\label{alg:PGFS_theoretical}
\begin{algorithmic}[1]
\REQUIRE Root node $x_0$; learned model $\mathcal M_\theta = (v_\theta, T_\theta,\gamma_\theta)$;
temperature $\alpha$;
number of simulations $N$.
\ENSURE Root policy $\pi'$ and non-stationary policy $\widetilde\pi$

\STATE Initialize frontier $\mathcal{F} \leftarrow \{x_0\}$
\STATE Initialize statistics:
$l(x_0)\gets 0, g(x_0)\gets0, d(x_0)\gets1, V(x_0)\gets v_\theta(x_0)$

\FOR{$h = 1$ to $N$}
    \STATE Sample node $x^\star \in \mathcal{F}$ according to
    \[
    P(x) \propto \exp\!\left(l(x) + \alpha  V(x)\right)
    \]
    
    \STATE Remove $x^\star$ from the frontier
    \[\mathcal F\gets \mathcal F - \{x^\star\}\]
    
    \FOR{\textbf{in parallel} $a\in\mathcal A$}
    \STATE Get child node $y$, reward $r$, prior probability $p$, and effective discount from the model
    \[y,r,p\gets T_\theta(x^\star, a),\qquad
    \gamma(x^\star,y)\gets\gamma_\theta(x^\star,a)\]
    \STATE Update the statistics for $y$
    \[g(y) \gets g(x^\star) + d(x^\star)\cdot r\]
    \[l(y)\gets l(x^\star) + \log p\]
    \[d(y) \gets d(x^\star)\cdot \gamma(x^\star,y)\]
    \[V(y) \gets g(y) + d(y) \cdot v_\theta(y)\]
    \STATE Add the child to the frontier $\mathcal F \gets \mathcal F\cup \{y\}$
    \ENDFOR

\ENDFOR

\STATE Compute the improved policy
\[
\pi'(a) \propto \sum_{x\in \mathcal F_a}\exp(l(x) + \alpha V(x)), \quad a\in \mathcal A
\]
where $\mathcal F_a$ is the part of the frontier in the subtree of the root action $a$.

\STATE Define $\widetilde\pi$: execute $a\sim\pi'$, sample $x\in\mathcal F_a$ with probability proportional to $\exp(l(x)+\alpha V(x))$, follow the remaining path to $x$, then follow the prior policy $\pi$.
\STATE \textbf{return} $\pi',\widetilde\pi$
\end{algorithmic}
\end{algorithm}

\subsection{Proof of the Theorems in \cref{subsubsec:theory}}
\label{appsubsec:proof}

We first introduce the following lemma.

\begin{lemma}[Donsker–Varadhan]
\label[lemma]{lemma:DV}
Suppose $\Omega$ to be a finite support of all the functions related here.

Let $Z(\alpha)=\sum_{x\in\Omega}P(x)e^{\alpha A(x)}$, and $\bar\eta_\alpha(x)=\frac{P(x)e^{\alpha A(x)}}{Z(\alpha)}$.

Whenever $\epsilon$ and $\alpha$ satisfies 

\begin{equation}
    \epsilon=\mathrm{KL}(\bar\eta_\alpha\|P) \label{DV-eps-def}
\end{equation}

The penalized optimization problem

\[
\argmax_{\mu\in \Delta(\Omega)}\left\{\sum _{x\in \Omega}\mu(x)A(x)-\frac 1 \alpha \text{KL}(\mu\| P)\right\}
\]

and the constrained optimization problem

\[
\argmax_{\mu\in \Delta (\Omega)}\sum_{x\in \Omega}\mu(x)A(x)\\
\text{s.t. } \text{KL}(\mu\|P)\le \epsilon
\]

have the same solution $\bar\eta_\alpha$.

\end{lemma}

\begin{proof}
    This is a known result called Donsker–Varadhan variational formula. See proof of Proposition 1 in \cite{DBLP:journals/corr/abs-1712-08650}
\end{proof}

We the prove \cref{thm:leafselection}.

\LeafSelectionThm*

\begin{proof}
    By \cref{lemma:DV}, we only need to show that $P_{\mathcal F}$ is a distribution, i.e.

    \[\sum_{x\in \mathcal F}P_{\mathcal F}(x) = 1\]

    This can be proved by induction.

    \paragraph{Initial step:} At the very beginning of the search (line 1-3 in \cref{alg:PGFS_theoretical}), we have $\mathcal F=\{x_0\}$, and $l(x_0)=0$.
    Then it is easy to obtain $\sum_{x\in \mathcal F}P_{\mathcal F}(x) = \exp(l(x_0))=1$.

    \paragraph{Induction step:} Suppose we have $\sum_{x\in \mathcal F}P_{\mathcal F}(x)=1$ currently.
    Consider expanding any node $x^\star\in\mathcal F$.
    Let $y_1,\cdots, y_{|\mathcal A|}$ be its children, and $p_1,\cdots,p_{|\mathcal A|}$ be the prior probabilities defined in line 10 of \cref{alg:PGFS_theoretical}.
    Then we have

    \begin{align*}
       & \sum_{u\in \mathcal F'}P_{\mathcal F'}(u)\\
       = &\sum_{u\in (\mathcal F-\{x^\star\})\cup \{y_i\}_{i=1}^{|\mathcal A|}}\exp(l(u))\\
       = &\sum_{u\in \mathcal F}\exp(l(u))-\exp(l(x^\star)) + \sum_{i=1}^{|\mathcal A|}\exp(l(y_i))\\
       = & \sum_{u\in \mathcal F}P_{\mathcal F}(u)-\exp(l(x^\star)) +\sum_{i=1}^{|\mathcal A|}\exp(l(x^\star)+\log p_i)\\
       = & 1-\exp(l(x^\star)) +\exp(l(x^\star))\sum_{i=1}^{|\mathcal A|}p_i\\
       = & 1
    \end{align*}
\end{proof}

Before we prove \cref{thm-informal:PI}, we introduce the following lemma:

\begin{lemma}

    Let $x_0$ be the root, $p(a):=P(a|x_0)$ is the prior of the root action $a$, $\mathcal F_a$ be the intersection of $\mathcal F$ and the subtree of root action $a$, and $\bar P_a(x) := P(x)/p(a)$ for $x\in \mathcal F_a$.
    
    Denote $\rho(a)\propto \sum _{x\in \mathcal F_a}\exp(l(x) + \alpha V(x))$ for all root actions as the probability we selection root action $a$ in the selection phase. Also denote $M_a = \max_{x\in \mathcal F_a}V(x)$ as the maximal path-based value in $\mathcal F_a$. 
    
    Then we have
    
    1.  $\rho = \mathrm{softmax}(G)$, where
    
    \[
    G(a) = \log p(a) + \alpha f(a),\quad
    f(a)=\frac 1 \alpha\log \sum_{x\in \mathcal F_a}\bar P_a(x)\exp(\alpha V(x))
    \]
    
    2. $M_a+\frac 1 \alpha \log p^*\le f(a)\le M_a$, where $p^* = \sum _{x\in \mathcal F_a, V(x)=M_a}\bar P_a(x)>0$.
    
    This implies $f(a)\to M_a$ when $\alpha\to \infty$ and $\rho\to \argmax_a M_a$ when $\alpha\to \infty$.
\end{lemma}

\begin{proof}
The first formula is straightforward.

\begin{align*}
\rho(a)&\propto \sum _{x\in \mathcal F_a}\exp(l(x) + \alpha V(x))\\
&\propto\exp\left(\log  \sum _{x\in \mathcal F_a}P(x)\exp(\alpha V(x))\right)\\
&=\exp\left(\log p(a) +\log \sum_{x\in\mathcal F_a}\bar P_a(x)\exp(\alpha V(x))\right)\\
&=\exp \left(\log p(a) + \alpha f(a)\right)\\
&=\exp(G(a))
\end{align*}

For the second formula, we have

\[
f(a) = \frac 1 \alpha\log \sum_{x\in \mathcal F_a}\bar P_a(x)\exp(\alpha V(x)) \le \frac 1 \alpha \log \exp(\alpha M_a) = M_a
\]

and

\begin{align*}
f(a)&=\frac 1 \alpha\log \sum_{x\in \mathcal F_a}\bar P_a(x)\exp(\alpha V(x))\\
&\ge \frac 1 \alpha\log \sum_{x\in \mathcal F_a,V(x)=M_a}\bar P_a(x)\exp(\alpha M_a)\\
&=\frac 1 \alpha\log\left(p^*\exp(\alpha M_a)\right)\\
&=M_a+\frac 1 \alpha \log p^*\\
\end{align*}

\end{proof}

We then restate and prove the formal version of \cref{thm-informal:PI}.
Let $q=\eta$ be the final frontier selection distribution. The returned policy $\widetilde\pi$ induces this same path distribution, since
\[
\pi'(a)q(x\mid a)
=\left(\sum_{y\in\mathcal F_a}q(y)\right)
\frac{q(x)}{\sum_{y\in\mathcal F_a}q(y)}
=q(x),\qquad x\in\mathcal F_a.
\]

\begin{proposition}
    Assume an accurate model of transitions, rewards, and discount factors, and the state-value prediction $v_\theta(x)$ accurately predicts $V^\pi(x)$, where $\pi$ is the prior policy, then for the non-stationary policy $\widetilde\pi$ returned by \cref{alg:PGFS_theoretical}, we have

    \[
    J(\widetilde\pi)=\sum_{x\in \mathcal F}q(x)V(x)\ge V^\pi(s)
    \]
    
    where $s$ is the root and $J(\widetilde\pi)$ denotes the expected return of executing $\widetilde\pi$ from $s$.
    
    This gives a non-stationary policy that improves the current policy $\pi$.
\end{proposition}

\begin{proof}
    If $v_\theta(x)=V^\pi(x)$, induction shows that, after the root has been expanded, the following identities hold at every search iteration:
    
    \begin{align*}
    V^\pi(s) &= \sum_{x\in \mathcal F}P(x)V(x),\\
    Q^\pi(s,a) &= \sum_{x\in \mathcal F_a}\bar P_a(x)V(x).
    \end{align*}
    
    The frontier is a cut of the tree, and each expansion preserves its prior-weighted value.
    Specifically, the path-statistic updates give $g(y)=g(x)+d(x)r(x,y)$ and $d(y)=d(x)\gamma(x,y)$ for each child $y$ of $x$.
    The Bellman equation for the exact value function, using the same effective edge discounts, therefore gives
    \begin{align*}
    \sum_{y\in\operatorname{sons}(x)}\pi(y\mid x)V(y)
    &=g(x)+d(x)\sum_{y\in\operatorname{sons}(x)}\pi(y\mid x)
    \bigl[r(x,y)+\gamma(x,y)v_\theta(y)\bigr]\\
    &=g(x)+d(x)v_\theta(x)=V(x).
    \end{align*}
    Together with $P(y)=P(x)\pi(y\mid x)$, this yields
    
    \[
    \forall x: P(x)V(x) = P(x)\sum_{y\in \text{sons}(x)}\pi(y|x)V(y) = \sum _{y\in \text{sons}(x)}P(y)V(y) 
    \]
    
    Thus we have

    \begin{align*}
    J(\widetilde\pi)&=\sum_{x\in \mathcal F}q(x)V(x)\\
    &\ge \sum _{x\in \mathcal F}q(x)V(x) - \frac 1 \alpha \mathrm{KL}(q\|P)\\
    &\ge \sum_{x\in \mathcal F}P(x)V(x)-\frac 1 \alpha \mathrm{KL}(P\|P)&\quad\text{\cref{thm:leafselection}}\\
    &=V^\pi(s)
    \end{align*}

\end{proof}

\subsection{Time Complexity}
\label{appsubsec:timecomplexityanalysis}

We briefly analyze the theoretical time complexity of MCTS and PGFS.
Here we denote $N$ as the number of simulations, $K = |\mathcal A|$ as the branching factor.
We also consider the time complexity of one model forward pass as $C$.

\paragraph{Time Complexity of MCTS}
The vanilla MCTS conducts $N$ simulations, each with three stages: Selection, Expansion, and Back Propagation.
The expansion stage requires a constant number of model forward passes, and therefore has time complexity $O(C)$.
Both selection and backpropagation traverse the tree from the root to a leaf.
In the worst case, the tree depth can be $\Theta(i)$ at the $i$-th simulation, so each of these two stages takes $O(i)$ time.
Therefore, the total time complexity over all simulations is
\[
O\left(\sum_{i=1}^N (C + i)\right) = O(C N + N^2).
\]
When executed on GPUs, model forward passes are highly parallelized and typically much faster than tree traversal.
Thus, when $C$ is treated as a constant with respect to $N$, the overall complexity is dominated by tree traversal and becomes $O(N^2)$.

\paragraph{Time Complexity of Gumbel MCTS}
The Gumbel MCTS uses a sequential halving procedure for root actions, which divides the whole $N$ simulations into $c =\lfloor \log_2 K\rfloor$ stages.
In each stage, Gumbel MCTS performs simulations evenly for the remaining subtrees, and drops half of them.
In this way, there are $\left\lfloor\frac{K}{2^{i-1}}\right\rfloor$ subtrees remaining in stage $i$, each with the same size
\[\sum_{j=1}^i\left\lfloor\frac{\left\lfloor\frac{N}{c}\right\rfloor}{\left\lfloor\frac{K}{2^{j-1}}\right\rfloor}\right\rfloor=\left\lfloor\frac{N}{c}\right\rfloor O\left(\sum_{j=1}^i\frac{2^{j-1}}{K}\right)=O\left(\frac{N2^i}{cK}\right)\]
And the total complexity for selection and backpropagation in the worst case is
\[\sum_{i=1}^c \left\lfloor\frac{K}{2^{i-1}}\right\rfloor O\left(\left(\frac{N2^i}{cK}\right)^2\right) = \sum_{i=1}^c O\left(\frac{N^22^i}{c^2K}\right)=O\left(\frac {N^2} {c^2}\right)=O\left(\frac {N^2} {\log^2K}\right)\]
The complexity for expansion is still $O(CN)$, making the total complexity $O\left(CN+ \frac {N^2}{\log^2K}\right)$.
When $C$ and $K$ are both treated as constants, the overall complexity is also $O(N^2)$.

\paragraph{Time Complexity of PGFS on a single CPU}
Consider one simulation of our PGFS (\cref{alg:PGFS_theoretical}).
The selection phase in $i$-th iteration is a sampling over the frontier of the size $iK$, which can be done in $O(\log i + \log K)$ with a Fenwick tree \citep{Fenwick1994AND}.
The expansion phase involves a network forward pass, one point update (removal of $x^\star$), and append-only insertions for the $K$ newly generated children.
The point update costs $O(\log i+\log K)$.
For the append-only insertions, we use the lazy construction of the Fenwick tree: when appending the $m$-th entry, we only construct the newly activated Fenwick node and do not update future nodes.
This costs $O(1+\nu_2(m))$, where $\nu_2(m)$ is the exponent of the largest power of two dividing $m$.
Since $\sum_{m=1}^M \nu_2(m)=O(M)$, each append-only insertion has amortized $O(1)$ cost.
Thus, inserting the $K$ newly generated children costs amortized $O(K)$ per simulation.
Therefore, the overall time complexity of PGFS on a single CPU core is
\[O\left(\sum_{i=1}^N(C + \log i+\log K + K)\right) = O(CN + N\log N + NK)\]

\paragraph{Time Complexity of PGFS on GPU}
PGFS is more suitable for GPUs, where abundant parallelism is available.
We assume enough GPU parallelism, so that parallelizable operation can be fully parallelized.
Then for PGFS (\cref{alg:PGFS_theoretical}), a sampling over the frontier can be easily implemented as a tensor operation with parallel-time complexity also in $O(\log i + \log K)$, which has a smaller constant in complexity if we implement segment tree on GPU.
Apart from the network forward pass, the expansion phase can be parallelized with $O(1)$ parallel time.
Therefore, the overall parallel-time complexity of PGFS is
\[O\left(\sum_{i=1}^N(C + \log i+\log K)\right) = O(CN + N\log N + N\log K)\]

\paragraph{Is the worst case realized in practice?}
The bounds above are worst-case complexities; a natural concern is whether this worst case is actually reached in practice, or whether it is a loose upper bound that rarely binds.
\cref{fig:search-scalability-k10} provides direct empirical evidence that it is reached: as the simulation budget $N$ grows, the measured runtime of MCTS and Gumbel MCTS scales quadratically, closely tracking the $O(N^2)$ bound, while PGFS scales quasi-linearly, consistent with $O(N\log N)$.
This confirms that the asymptotic gap between PGFS and MCTS-style search is not merely a theoretical artifact of worst-case analysis, but is realized in practice.

\clearpage
\section{Hyper-Parameters}
\label{appsec:hyperparam}

Our model structure follows MAZero \citep{liu2024efficient}, except using an additional head $G$ to predict the discount factor.

$$\mathbb P(\gamma_t = 1|s_t) = G_\theta(\mathbf {s}, \mathbf a)$$

This head is trained with BCE Loss.

Besides, since the improved policy produced by PGFS already incorporates value information.
We do not use AWPO here for policy training, instead, we use a simple Cross Entropy Loss for policy training.

To address the exponentially large joint action space in MARL, we follow MAZero \citep{liu2024efficient} and Sampled MuZero \citep{hubert2021learningplanningcomplexaction} and apply standard action down-sampling: at each node, we sample $K$ joint actions and restrict search to this subset.

The GPU MCTS and GPU Gumbel MCTS baselines used in our ablation studies are implemented based on DeepMind's open-source JAX library \texttt{mctx} \citep{deepmind2020jax}.

The base configuration of our method is shown in \cref{tab:hyperparam}.

\begin{table}[htb]
\centering
\caption{Hyperparameter settings.}
\label{tab:hyperparam}
\begin{tabular}{l c}
\toprule
\textbf{Parameter} & \textbf{Setting} \\
\midrule
Batch size              & $256$ \\
Buffer size             & $10^6$ \\
Learning rate     & $5 \times 10^{-4}$ \\
Gradient clipping       & $10$ \\
$\epsilon$-greedy       & $1.0 \rightarrow 0.05$ \\
$\epsilon$ annealing time & $10000$ \\
Action down-sampling $K$ & $40$\\
PGFS simulations $N$ & $100$\\
$\alpha$ in frontier selection & $200$\\
Number of Envs& $8$\\
Model update frequency (per Env Step) & $0.5$\\
Reward loss coefficient & $1$\\
Value loss coefficient & $0.25$\\
Policy loss coefficient & $1$\\
Consistency loss coefficient & $2$\\
Discount loss coefficient & $1$\\
Prior entropy regularization loss & $0.25$\\
Action selection type & Greedy \\
fc\_discount\_layers & $[32]$ \\
fc\_dynamic\_layers & $[128,128]$ \\
fc\_policy\_layers & $[32]$ \\ 
fc\_representation\_layers & $[128,128]$ \\
fc\_reward\_layers & $32$ \\
fc\_value\_layers & $32$ \\
Hidden state size & $128$ \\
Number of rollout steps & $5$ \\
TD steps & $5$ \\
Observation n\_stack & $1$ \\
pred\_hid & $64$ \\
pred\_out & $128$ \\
proj\_hid & $128$ \\
proj\_out & $128$ \\
prioritized\_replay\_eps & $10^{-6}$ \\
priority\_prob\_alpha & $0.6$ \\
priority\_prob\_beta & $0.4$ \\ 
Reward support & $[-2,2,5]$ \\
root\_dirichlet\_alpha & $0.3$ \\ 
root\_exploration\_fraction & $0.25$ \\
\bottomrule
\end{tabular}

\end{table}

We use the base configuration for all SMAX maps.

For MPE and OverCooked tasks, we enable \texttt{Model predict discount} as \texttt{True} to handle the reset of the environment after fixed steps.
We also enable discount prediction for MAZero and MAMBA on these tasks.

\clearpage
\section{Baselines}

\begin{figure*}[h]
\centering
\includegraphics[width=0.8\textwidth]{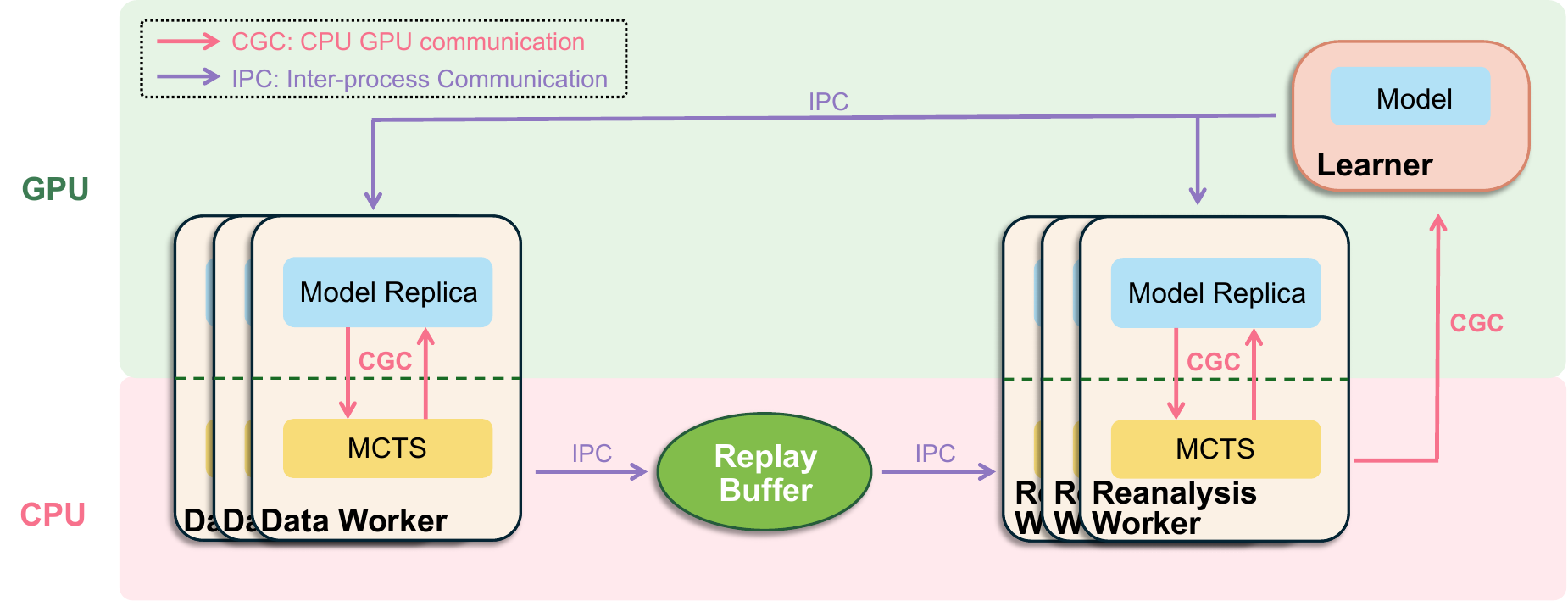}
\caption{Overview of MAZero's CPU Centric Architecture.}
\label{fig:MAZero-architecture}
\end{figure*}

We include the CPU-centric architecture of MAZero here for comparison.
As shown in \cref{fig:MAZero-architecture}, MAZero primarily relies on CPU-side orchestration for coordinating simulation, inference, and training, inducing frequent CPU-GPU communication which serves as a contrast to our GPU-centric architecture (\cref{fig:MATCHA-architecture}).

\end{document}